\documentclass[a4paper,11pt]{article}

\usepackage{amssymb,amsmath}

\newcommand{\ns}{\normalsize}
\newcommand{\be}{\begin{equation}}
\newcommand{\ee}{\end{equation}}
\newcommand{\ba}{\begin{eqnarray}}
\newcommand{\ea}{\end{eqnarray}}

\newcommand{\la}{\lambda}
\newcommand{\si}{\sigma}

\newcommand{\B}{\bar}
\newcommand{\ti}{\tilde}
\newcommand{\ph}{\phantom}
\newcommand{\nn}{\nonumber}
\newcommand{\ul}{\underline}

\numberwithin{equation}{section}

\begin{document}


\begin{titlepage}

\vspace{-3cm}

\title{\hfill{\ns hep-th/0606285\\}
   \vskip 1cm
   {\Large Four-dimensional Effective M-theory on a Singular $G_2$ Manifold}\\}
   \setcounter{footnote}{0}
\author{
{\ns\large Lara B Anderson$^1$\footnote{email: anderson@maths.ox.ac.uk},
 Adam B Barrett$^2$\footnote{email: barrett@thphys.ox.ac.uk},
Andr\'e Lukas$^2$\footnote{email: lukas@physics.ox.ac.uk}
\setcounter{footnote}{3}
and Masahiro Yamaguchi$^3$\footnote{email: yama@tuhep.phys.tohoku.ac.jp}} \\[1em]
   {\ns\it$^1$Mathematical Institute, University of Oxford}\\
   {\ns 24-29 St.~Giles', Oxford OX1 3LB, UK}\\[1em]
   {\it\ns $^2$The Rudolf Peierls Centre for Theoretical Physics}\\
   {\it\ns University of Oxford}\\
   {\ns 1 Keble Road, Oxford OX1 3NP, UK}\\[1em]
   {\it\ns $^3$Department of Physics, Tohoku University}\\
   {\ns Sendai, 980-8578, Japan}}
\date{}

\maketitle

\begin{abstract}\noindent
  We reduce M-theory on a $G_2$ orbifold with co-dimension four
  singularities, taking explicitly into account the additional gauge
  fields at the singularities. As a starting point, we use
  11-dimensional supergravity coupled to seven-dimensional
  super-Yang-Mills theory, as derived in a previous paper. The
  resulting four-dimensional theory has $\mathcal{N}=1$ supersymmetry
  with non-Abelian $\mathcal{N}=4$ gauge theory sub-sectors. We
  present explicit formulae for the K\"ahler potential, gauge-kinetic
  function and superpotential. In the four-dimensional theory,
  blowing-up of the orbifold is described by a Higgs effect induced by
  continuation along D-flat directions. Using this interpretation, we
  show that our results are consistent with the corresponding ones
  obtained for smooth $G_2$ spaces. In addition, we consider the
  effects of switching on flux and Wilson lines on singular loci of
  the $G_2$ space, and we discuss the relation to $\mathcal{N}=4$ SYM
  theory.
\end{abstract}

\thispagestyle{empty}

\end{titlepage}


\section{Introduction}

As is well known, the low-energy approximation for M-theory on a
smooth space-time is 11-dimensional supergravity
\cite{Witten:1995ex}. On a singular space-time this approximation is
also valid in the smooth regions. However, there is, in general, an
extra field content localized at each singularity. For example,
M-theory on $\mathbb{R}^{1,9}\times S^1 / \mathbb{Z}_2$ is
approximated by the famous Ho\v rava-Witten theory \cite{Hor-Wit} at
low-energies. In this case, 11-dimensional supergravity is coupled
to two 10-dimensional $E_8$ super-Yang-Mills theories, one living at
each of the orbifold fixed planes.

The space-times we consider in this paper contain co-dimension four
singularities of A-type, that is, singularities of the form
$\mathbb{C}^2/\mathbb{Z}_N$. It is well-known~\cite{Acharya,Gukov}
that the additional states which appear at these singularities are
$\mathrm{SU}(N)$ gauge multiplets.  The associated effective theory
for M-theory on spaces of the form
$\mathcal{M}_7\times\mathbb{C}^2/\mathbb{Z}_N$, where $\mathcal{M}_7$ is a smooth
seven-dimensional space with Minkowski signature, has been derived for
the first time in a recent paper~\cite{11-7}. This theory couples
11-dimensional supergravity, with fields suitably constrained in
accordance with the orbifold symmetry, to seven-dimensional
$\mathcal{N}=1$ super-Yang-Mills at the orbifold fixed point. With
this action available, it is now possible to carry out an explicit
reduction to four-dimensions of M-theory on a space containing
$\mathbb{C}^2/\mathbb{Z}_N$ singularities. These singularities are of
particular importance in the context of M-theory on $G_2$ spaces. As
is well-known~\cite{Comp2}, M-theory on a smooth $G_2$
space~\cite{joyce1}--\cite{deCarlos:2004ci} leads to
phenomenologically uninteresting four-dimensional $\mathcal{N}=1$
theories with Abelian vector multiplets and uncharged chiral
multiplets. For this reason, singular $G_2$ spaces have been
considered in the
literature~\cite{Acharya,Gukov},\cite{Acharya:2000gb}--\cite{Metzger}. In
particular, compactification with A-type singularities, which lead to
four-dimensional $\mathrm{SU}(N)$ gauge multiplets, can be seen as a
first step towards successful particle phenomenology from
M-theory on $G_2$ spaces.\\

The main theme of this paper is to study such compactifications of
M-theory, using the action of Ref.~\cite{11-7} to explicitly describe
the non-Abelian gauge fields which arise at the singularities.  The
relevant $G_2$ spaces are constructed by dividing a seven-torus $\mathcal{T}^7$
by a discrete symmetry group $\Gamma$, such that the resulting
singularities are of co-dimension four and of A-type. We will then
perform the reduction to four dimensions on these spaces. This
includes the reduction of the seven-dimensional $\mathrm{SU}(N)$ gauge
theories on the three-dimensional singular loci within the $G_2$
space. For the orbifold examples considered in this paper, the
singular loci will always be three-tori, $\mathcal{T}^3$. Hence, while the full
four-dimensional theory is $\mathcal{N}=1$ supersymmetric, the gauge
sub-sectors associated to each singularity have enhanced
$\mathcal{N}=4$ supersymmetry. Let us split the four-dimensional field
content into ``bulk fields'' which descend from 11-dimensional
supergravity and ``matter fields'' which descend from the
seven-dimensional super-Yang-Mills theories at the singularities. The
bulk fi§elds correspond to the moduli of the $G_2$ space (plus
associated axions from the M-theory three-form). For the $G_2$ spaces
we consider, the only geometrical parameters that survive the
orbifolding are the seven radii of the torus. Hence we have seven
moduli $\tilde{T}^A$ in the reduced theory. At an orbifold singularity these
moduli can be divided up into $\tilde{T}^0$ and $\tilde{T}^{mi}$, where $m=1,2,3$ and
$i=1,2$. The modulus $\tilde{T}^0$ can be thought of as corresponding to the
volume of the three-torus locus of the singularity, while the $\tilde{T}^{mi}$
depend on one radius $R^m$ of the singular locus and two radii of $\mathcal{T}^7$
transverse to the singularity. For each singularity, the matter fields
consist of $\mathcal{N}=1$ vector multiplets with gauge group
$\mathrm{SU}(N)$, plus three chiral multiplets $\mathcal{C}^{am}$ per
vector multiplet, where $m=1,2,3$ and $a$ is an adjoint index of the
gauge group. This is indeed the field content of the $\mathcal{N}=4$
theory in $\mathcal{N}=1$ language. \\

Let us now summarise our main results. If we focus on one particular singularity,
the K\"ahler potential, gauge-kinetic function and superpotential of the
four-dimensional effective theory are given by
\ba
K&=& \frac{7}{\kappa_4^2}\ln 2 - \frac{1}{\kappa_4^2}\sum_{A=0}^6 \ln (\tilde{T}^A
+ {\bar{\ti{T}}}^A ) +
\frac{1}{4\lambda_4^2}\sum_{m=1}^3\frac{(\mathcal{C}_a^m+\bar{\mathcal{C}}_a^m)
(\mathcal{C}^{am}+\bar{\mathcal{C}}^{am})}{(\tilde{T}^{m1}+\bar{\tilde{T}}^{m1})
(\tilde{T}^{m2}+\bar{\tilde{T}}^{m2})}\,, \label{kahlerintro}\\
f_{ab}&=&\frac{1}{\lambda_4^2}\tilde{T}^0\delta_{ab}\, , \label{gaugeintro} \\ W
&=& \frac{\kappa_4^2}{24\lambda_4^2}f_{abc}\sum_{m,n,p=1}^3
\epsilon_{mnp}\mathcal{C}^{am}\mathcal{C}^{bn}\mathcal{C}^{cp}\,. \label{supotintro}
\ea
These expressions constitute the leading order terms of series
expanded in terms of the parameter $h_4=\kappa_4/\lambda_4$, that is, the
gravitational coupling divided by the gauge coupling. Complete
expressions for the above quantities involve a sum over all
singularities and are given in Section~\ref{reduction}.

It is interesting to compare these results with those found by
compactification on the associated smooth $G_2$ space, obtained by
blowing-up the singularities. The construction of smooth $G_2$ spaces
in this way has been pioneered by Joyce~\cite{Joyce} and the
associated four-dimensional effective theories for M-theory on such
$G_2$ spaces have been computed in Refs.~\cite{Lukas,Barrett}.  For an
explicit comparison, it is useful to recall that the geometrical
procedure of blowing-up can be described within the $\mathrm{SU}(N)$ gauge
theory as a Higgs effect, whereby VEVs are assigned to the (real parts
of the) Abelian matter fields $\mathcal{C}^{im}$ along the D-flat
directions~\cite{11-7}.  This generically breaks $\mathrm{SU}(N)$ to its
maximal Abelian subgroup $\mathrm{U}(1)^{N-1}$ and leaves only the $3(N-1)$ chiral
multiplets $\mathcal{C}^{im}$ massless. This field content corresponds
precisely to the zero modes of M-theory on the blown-up geometry, with
the Abelian gauge fields arising as zero modes of the M-theory
three-form on the $N-1$ two-spheres of the blow-up and the chiral
multiplets corresponding to its moduli.  With this interpretation of
the blowing-up procedure we can compare the above K\"ahler potential,
restricted to the Abelian matter fields $\mathcal{C}^{im}$, to
the smooth result. We find that subject to a suitable embedding of
the Abelian group $\mathrm{U}(1)^{N-1}$ into $\mathrm{SU}(N)$ they are indeed exactly
the same, provided the Abelian matter fields $\mathcal{C}^{im}$ are
identified with the blow-up moduli. Also note that, when restricted
to the Abelian fields $\mathcal{C}^{im}$, the above superpotential~\eqref{supotintro}
vanishes (consistent with the result one obtains in the smooth case).

We  also  show that  the  superpotential~\eqref{supotintro}  can  be
obtained from a Gukov-type formula which involves the integration
of  the  complexified   Chern-Simons  form  of  the  seven-dimensional
$\mathrm{SU}(N)$  gauge theory  over the  internal three-torus  $\mathcal{T}^3$.  In
addition, we show explicitly  that the superpotential for Abelian flux
of the $\mathrm{SU}(N)$  gauge fields on $\mathcal{T}^3$ is correctly obtained from
the same  Gukov formula, an observation first made in Ref.~\cite{AcharyaMod}.  This
result  provides a further  confirmation for the matching  between the
singular  and smooth theories  as the flux  superpotential in  both limits
is of the same form if the field identification suggested by the
comparison of K\"ahler potentials is used.

Finally, we consider one of the $\mathcal{N}=4$ sectors of our action with gravity
switched off. We point out that singular and blown-up geometries correspond
to the conformal phase and the Coulomb phase of this  $\mathcal{N}=4$ theory, respectively.
Additionally, its S-duality symmetry translates into a T-duality on the singular $\mathcal{T}^3$ locus.
We speculate about a possible extension of this S-duality to the full supergravity
theory and analyse some of the resulting consequences. \\

The plan of the paper is as follows. In Section~\ref{review} we write
down the effective supergravity action for M-theory on a space-time
$\mathcal{M}_7\times\mathbb{C}^2/\mathbb{Z}_N$.  The reduction of this theory on
singular $G_2$ spaces is then carried out in Section~\ref{reduction},
and we discuss the identification of the results with those from
reduction on a smooth manifold. In Section~\ref{wilflux}, we study the
effect of more complicated backgrounds, with non-vanishing Wilson
lines and flux and also discuss the relation to $\mathcal{N}=4$ SYM
theory. In addition, there are two appendices containing technical
details. In the first appendix, the construction, geometry, and
topology of orbifold based $G_2$ manifolds is discussed. There is also
a list of sixteen possible orbifold groups that lead to singular $G_2$
manifolds for which the results of this paper are directly
applicable. The second appendix reviews the reducing M-theory on a
smooth $G_2$ manifold, and then presents the K\"ahler potential for
M-theory on the blown-up orbifolds.  \\

Let us state the index conventions we shall use for the various spaces we
consider. We take 11-dimensional space-time to have mostly positive
signature, that is $(-++\ldots+)$, and use indices
$M,N,\ldots=0,1,\ldots,10$ to label 11-dimensional coordinates
$(x^M)$.  Four-dimensional coordinates on $\mathbb{R}^{1,3}$ are
labelled by $(x^\mu)$, where $\mu,\nu,\ldots=0,1,2,\ldots,3$, while
points on the internal $G_2$ space $\mathcal{Y}$ are labelled by
coordinates $(x^A)$, where $A,B,\ldots = 4,\ldots ,10$.  The
coordinates of co-dimension four singularities in the internal space
will be denoted by $(y^{\hat{A}})$, $\hat{A},\hat{B}\ldots=7,8,9,10$,
while the complementary three-dimensional singular $\mathcal{T}^3$ locus has
coordinates $x^m$, with $m,n,\ldots =4,5,6$. To describe the
seven-dimensional gauge theories it is also useful to introduce
coordinates $(x^{\hat{\mu}})$,
$\hat{\mu},\hat{\nu},\ldots=0,1,\ldots,6$ on the locus of the
singularity. Underlined versions of all the above index types denote
the associated tangent space indices. We frequently use the term
``bulk'' to refer to the full 11-dimensional space-time.


\section{M-theory on A-type singularities} \label{review}
In this section we review the effective supergravity action for M-theory on
the space $\mathcal{M}_{11}=\mathcal{M}_7\times \mathbb{C}^2/\mathbb{Z}_N$, as constructed
in Ref.~\cite{11-7}. This space has a seven-dimensional singular locus
at the origin of $\mathbb{C}^2$ on which an $\mathcal{N}=1$ super-Yang-Mills (SYM)
theory with gauge group $\mathrm{SU}(N)$ resides. The construction of Ref.~\cite{11-7} couples 11-dimensional
supergravity to this seven-dimensional SYM theory in a locally supersymmetric way.
Hence the basic structure of this theory is
\begin{equation}
 \mathcal{S} = \mathcal{S}_{11}+\mathcal{S}_7\; , \label{S11-7}
\end{equation}
with $\mathcal{S}_{11}$ being the action of 11-dimensional supergravity,
with fields suitably restricted in accordance with the orbifold symmetry, and
$\mathcal{S}_7$ being the seven-dimensional SYM action. This form of the
action is valid for a single singularity of the type $\mathbb{C}^2/\mathbb{Z}_N$.
Later, in the context of $G_2$ compactifications, we will need a trivial
extension of this action incorporating a number of similar seven-dimensional
actions $\mathcal{S}_7$, one for each singularity. In this
section, we will focus on a single singularity for simplicity. The above theory has
been constructed as an expansion in $h_7=\kappa_7/\la_7$, where
$\kappa_7=\kappa_{11}^{5/9}$ and $\la_7$ is the Yang-Mills coupling.
The coupling, $\lambda_7$, has been determined in Ref. \cite{Friedmann} and is given in terms of the
11-dimensional Newton constant $\kappa_{11}$ as
\begin{equation} \label{couplerel}
\la_7^2=(4\pi)^{4/3}\kappa_{11}^{2/3}\, .
\end{equation}
The bulk action, 11-dimensional supergravity, appears at zeroth order
in this expansion, while the Yang-Mills theory arises at higher order.
We now proceed to discuss the action~\eqref{S11-7} order by order in this
expansion, focusing on the bosonic parts. \\

The bosonic part of the 11-dimensional supergravity action \cite{Julia} is given by
\begin{equation} \label{11dsugra}
\mathcal{S}_{11}  =   \frac{1}{2\kappa_{11}^2}\int_{\mathcal{M}_{11}}
 \left( \hat{R}\ast \boldsymbol{1} -\frac{1}{2}G\wedge \ast G
 -\frac{1}{6}C\wedge G\wedge G \right)\, ,
\end{equation}
where $\hat{R}$ is the Ricci scalar, associated with the 11-dimensional metric
$\hat{g}$ and vielbein $\hat{e}$, and $C$ is a three-form field with field
strength $G=\mathrm{d}C$. The supersymmetric transformation of the gravitino
$\Psi$ is given by
\begin{equation} \label{susyPsi}
\delta \Psi_M  =  2\nabla_M\eta+\frac{1}{144}\left( \Gamma_M^{\phantom{M}NPQR}
 -8\delta_M^N\Gamma^{PQR}\right)\eta G_{NPQR}
\end{equation}
where $\eta$ is an 11-dimensional Majorana spinor and $\nabla_M$ is the spinor
covariant derivative defined by
\begin{equation}
\nabla_M=\partial_M+\frac{1}{4}\hat{\omega}_M^{\phantom{M}\underline{M}\underline{N}}
\Gamma_{\underline{M}\underline{N}}\, .
\end{equation}
We are working in the ``up-stairs'' picture for the orbifold and, hence,
the fields in this action have to be constrained by the orbifold symmetry.
The $\mathbb{Z}_N$ generator $R$, acting on the orbifold coordinates as
$y^{\hat{A}} \to {R^{\hat{A}}}_{\hat{B}} y^{\hat{B}}$, is explicitly taken to be
\begin{equation}
({R^{\hat{A}}}_{\hat{B}})=\left( \begin{array}{cccc}
\cos(2\pi/N) & -\sin(2\pi/N) & 0 & 0 \\
\sin(2\pi/N) & \cos(2\pi/N) & 0 & 0 \\
0 & 0 & \cos(2\pi/N) & -\sin(2\pi/N) \\
0 & 0 & \sin(2\pi/N) & \cos(2\pi/N)
\end{array} \right)\, .
\label{R}
\end{equation}
As an example of the $\mathbb{Z}_N$ restrictions which have to be imposed on
the fields we present the constraint on the purely internal part of the metric.
It is given by
\begin{equation}
 \hat{g}_{\hat{A}\hat{B}}(\hat{x},R\hat{y})={{\left(R^{-1}\right)}_{\hat{A}}}^{\hat{C}}
{{\left(R^{-1}\right)}_{\hat{B}}}^{\hat{D}}\hat{g}_{\hat{C}\hat{D}}(\hat{x},\hat{y})\; , \label{gcons}
\end{equation}
and analogously for the other field components. A full list of these constraints
can be found in Ref.~\cite{11-7}.

When the bulk fields are constrained to the orbifold point, $y=0$,
the above field restrictions turn into truncations of the fields. In
the language of seven-dimensional $\mathcal{N}=1$ supergravity,
these truncated fields then correspond to a gravity multiplet and
one (three) Abelian vector multiplet(s) for $\mathbb{Z}_N$ with
$N>2$ (for $\mathbb{Z}_2$). The various scalar fields in those
multiplets parameterise an $\mathrm{SO}(3,1)/\mathrm{SO}(3)$ coset
for $N>2$ and an $\mathrm{SO}(3,3)/\mathrm{SO}(3)^2$ coset for
$N=2$. The seven-dimensional part of the action which we are going
to discuss describes the coupling of these ``bulk-induced''
multiplets to the genuine seven-dimensional $\mathrm{SU}(N)$
multiplets. \\

We now proceed with a discussion, following Ref.~\cite{11-7}, of the
seven-dimensional super-Yang-Mills theory located at the seven-dimensional
singular locus $y=0$. It can be expanded as
\begin{equation} \label{SYMexp}
\mathcal{S}_7=\la_7^{-2}\left( \mathcal{S}_7^{(0)}+h_7^2\mathcal{S}_7^{(2)}+h_7^4\mathcal{S}_7^{(4)}+\ldots\right),
\end{equation}
where the $\mathcal{S}_7^{(n)}$, $n=0,2,4,\ldots$ are independent of $h_7$. The
leading order contribution $\mathcal{S}_7^{(0)}$ has been derived in
Ref.~\cite{11-7}, using results~\cite{Park,Berghshoeff} for seven-dimensional
$\mathcal{N}=1$ Einstein-Yang-Mills supergravity. Dropping the fermionic
terms from the result in Ref.~\cite{11-7}, one finds for the leading term
\ba \label{boseaction}
\mathcal{S}_{7}^{(0)}&=&  \int_{\mathbb{R}^{1,3}\times B}
     \mathrm{d}^7x \sqrt{-\hat{g}}\left(
     -\frac{1}{4}H_{ab}F^a_{\hat{\mu}\hat{\nu}}F^{b\hat{\mu}\hat{\nu}}
     -\frac{1}{2}H_{aI}F^a_{\hat{\mu}\hat{\nu}}F^{I\hat{\mu}\hat{\nu}}
     -\frac{1}{4}(\delta H)_{IJ}F^I_{\hat{\mu}\hat{\nu}}F^{J\hat{\mu}\hat{\nu}}      \right. \nn \\
&&\hspace{4.0cm} \left. -\frac{1}{2}e^\tau\hat{\mathcal{D}}_{\hat{\mu}}
     \phi_{au}\hat{\mathcal{D}}^{\hat{\mu}}\phi^{au}-\frac{1}{2}\left( \delta K      \right)_{I u J v}\partial_{\hat{\mu}}\ell^{Iu}\partial^{\hat{\mu}}\ell^{Jv}
     +\frac{1}{4}D^{au}D_{au}\right) \nn \\
&&-\frac{1}{4}\int_{\mathbb{R}^{1,3}\times B} C\wedge F^a \wedge F_a
     \, ,
\ea
where
\ba
H_{ab}&=&\delta_{ab}\, , \label{H1} \\
H_{aI}&=&2\,\phi_{au}{\ell_{I}}^u\, , \label{H2} \\
(\delta H)_{IJ}&=&2\,\phi_{au}{\phi^a}_v\ell_{I}^{\ph{I}u}\ell_{J}^{\ph{J}v}\, ,
     \label{H3} \\
\left( \delta K \right)_{I u J v}&=&e^\tau  \phi_{au}\phi^{a}_{\ph{a}v}
     \delta_{\alpha\beta}{\ell_{I}}^\alpha{\ell_J}^\beta\, , \\
D_{au}&=&\frac{i}{\sqrt{2}}\epsilon_{uvw}e^\tau f_{abc}\phi^{bv}\phi^{cw}\, .
\ea
The covariant derivatives that appear are given by
\ba
\mathcal{D}_{\hat{\mu}} A_{\hat{\nu} a}&=&\partial_{\hat{\mu}} A_{\hat{\nu} a}
     -\hat{\Gamma}^{\hat{\rho}}_{\hat{\mu}\hat{\nu}}A_{\hat{\rho} a}+
     f_{ab}^{\phantom{ab}c}A_{\hat{\mu}}^bA_{\hat{\nu}}^c\, , \\
\hat{\mathcal{D}}_{\hat{\mu}}\phi_{a}^{\ph{a}u}&=&\partial_{\hat{\mu}}
     \phi_{a}^{\ph{a}u} - \left( \ell^{Iu}\partial_{\hat{\mu}}
     \ell_{Iv}\right) {\phi_{a}}^v+{f_{ab}}^cA_{\hat{\mu}}^b{\phi_c}^u\, .
\ea

The bosonic fields localized on the orbifold plane are the
$\mathrm{SU}(N)$ gauge vectors $F^a=\mathcal{D}A^a$ and the
$\mathrm{SO}(3)$ triplets of scalars ${\phi_a}^u$, where $u=1,2,3$.
All other fields are projected from the bulk onto the orbifold plane
and correspond to the bosonic parts of the seven-dimensional gravity
and Abelian vector multiplets mentioned earlier. There are algebraic
equations relating them to the 11-dimensional fields in
$\mathcal{S}_{11}$. These relations are trivial for the metric
$\hat{g}_{\hat{\mu}\hat{\nu}}$ and three-form
$C_{\hat{\mu}\hat{\nu}\hat{\rho}}$, both part of the
seven-dimensional gravity multiplet. For the other fields we have
\begin{eqnarray}
\tau&=&\frac{1}{2}\ln\det \hat{g}_{\hat{A}\hat{B}} \, , \label{id1} \\
\ell_I^{\ph{I}\underline{J}}&=&\frac{1}{4}\mathrm{tr}\left( \B{t}_I v t^J v^{T}
   \right)\, ,\\
F^I_{\hat{\mu}\hat{\nu}}&=&-\frac{1}{4}\mathrm{tr}\left(
     t^IG_{\hat{\mu}\hat{\nu}}\right)\, . \label{idn}
\end{eqnarray}
Here $G_{\hat{\mu}\hat{\nu}}\equiv (G_{\hat{\mu}\hat{\nu}
\hat{A}\hat{B}})$, $v\equiv
(e^{\tau/4}\hat{e}^{\hat{A}}_{\ph{A}\underline{\hat{B}}})$, and
$\ell_I^{\ph{I}\underline{J}}\equiv ({\ell_I}^u,{\ell_I}^\alpha)$,
$\alpha=4,5,6$. The indices $I,J,\ldots=1,\ldots,6$ and the matrices
$t^I$ are the generators of $\mathrm{SO}(4)$. (For convenience they
are taken in a particular representation given in Ref.~\cite{11-7}.)
The scalar field $\tau$, effectively the overall scale factor of the
orbifold, is also part of the seven-dimensional gravity multiplet.
The matrices ${\ell_I}^{\underline{J}}$ contain the moduli of the
orbifold and parameterise an $\mathrm{SO}(3,1)/\mathrm{SO}(3)$ coset
for $N>2$ and an $\mathrm{SO}(3,3)/\mathrm{SO}(3)^2$ coset for
$N=2$. From the viewpoint of seven-dimensional supergravity, these
cosets contain the scalar fields in the Abelian vector multiplets.
Finally, the vector fields $F^I$ descend from the M-theory four-form
$G$, with three of them becoming part the seven-dimensional gravity
multiplet and the remaining ones accounting for the vector
multiplets. For the $\mathbb{Z}_2$ case, the orbifold symmetry
imposes no constraints on the $F^I$ and there are three
$\mathrm{U}(1)$ vector multiplets. For the case $N>2$, there is an
orbifold constraint, and this imposes $F^5=F^6=0$, so there is only
one $\mathrm{U}(1)$ vector multiplet present.  \\

We finish this section by giving the structure of fermionic
supersymmetry transformations at the singularity. Firstly, the gravitino sees a correction to its
supersymmetry transformation that, to leading order, is given schematically by
\begin{equation} \label{susyPsi2}
\delta \Psi \sim \frac{\kappa_{11}^2}{\lambda_7^2} \delta^{(4)}(y^{\hat{A}}) \left\{ \phi_a\hat{\mathcal{D}}\phi^a + e^\tau \left( F^I \ell_I\phi^a\phi_a +F^a\phi_a \right) + f_{abc}\phi^a\phi^b\phi^c \right\} \eta.
\end{equation}
Secondly, the leading order
supersymmetry transformation of the gaugino $\la^a$ takes the form
\begin{equation} \label{susyla}
\delta \la^a \sim \left\{ \hat{\mathcal{D}}\phi^a + e^\tau \left( F^I\ell_I\phi^a + F^a \right) + {f^a}_{bc}\phi^b\phi^c\right\} \eta \, .
\end{equation}
For the purposes of this paper precise formulae for these transformations are not needed. Such formulae are however computed in Ref.~\cite{11-7}, taking account of some subtleties with identification of 11- and 7-dimensional fermions.


\section{The four-dimensional effective action on a $G_2$  orbifold} \label{reduction}
In this section, we will calculate the four-dimensional effective
theory for M-theory on a $G_2$ orbifold, using the
action~\eqref{S11-7}, \eqref{11dsugra}, \eqref{boseaction} for
11-dimensional supergravity coupled to seven-dimensional
super-Yang-Mills theory. We assume the $G_2$ orbifold $\mathcal{Y}$ takes the
form $\mathcal{T}^7/\Gamma$, where $\mathcal{T}^7$ is a seven-torus and $\Gamma$ is a
discrete group of symmetries of $\mathcal{T}^7$. We assume further that the
fixed points of the orbifold group $\Gamma$ are all of co-dimension
four, and also that points on the torus that are fixed by one
generator of $\Gamma$ are not fixed by other generators. A class of
such orbifolds was constructed in Ref.~\cite{joyce1,joyce2,Joyce,Barrett}, and are
described in Appendix~\ref{A}. According to the ADE classification,
the singularities of $\mathcal{Y}$ are then all of type $A_{N-1}=\mathbb{Z}_N$,
for some $N$ \cite{Barrett}. Furthermore, the approximate form of $\mathcal{Y}$
near a singularity is $\mathbb{C}^2/\mathbb{Z}_{N}\times \mathcal{T}^3$, where
$\mathcal{T}^3$ is a three-torus. The coordinates of the underlying torus $\mathcal{T}^7$
are denoted by $(x^A)$, where we change the range of indices to
be $A,B,\ldots = 1,\ldots ,7$, for convenience. For most of this
section, we will focus on one such singularity for simplicity. In the
neighbourhood of a singularity, M-theory is described by the
seven-dimensional action reviewed in the previous section. Without
loss of generality, we consider this singularity of $\mathcal{Y}$ to be located
at $x^4=x^5=x^6=x^7=0$ and we split coordinates according to
\begin{equation} \label{7coordid}
(x^A)\to (x^m,y^{\hat{A}})\; ,
\end{equation}
where $m,n,\ldots =1,2,3$ and $\hat{A},\hat{B},\ldots = 4,5,6,7$.
The generator~\eqref{R} of the $\mathbb{Z}_N$ symmetry then acts
on the coordinates $(y^{\hat{A}})$.

For the purpose of our calculation, it is convenient to work with the orbifold
rather than any partially blown-up version thereof. However, we will see
later that the possibility of blowing up some of the singularities can,
in fact, be effectively described within the low-energy four-dimensional
gauge theories associated with the singularities.

\subsection{Background solution and zero modes}
Let us now discuss the M-theory background on
$\mathcal{M}_{11}=\mathbb{R}^{1,3}\times \mathcal{Y}$. Throughout our
calculations we take the expectation values of fermions to vanish. This means
that we only need concern ourselves with the bosonic equations of motion. For
the bulk 11-dimensional supergravity these are
\ba
\mathrm{d}G&=&0\, , \label{bulkeom1} \\
\mathrm{d}\ast G &=& -\frac{1}{2}G\wedge G\, ,  \label{bulkeom2}\\
\hat{R}_{MN}&=&\frac{1}{12}\left( G_{MPQR}{G_N}^{PQR}-\frac{1}{12}\hat{g}_{MN}G_{PQRS}G^{PQRS} \right)\, . \label{bulkeomn}
\ea
We take the background metric $\langle\hat{g}\rangle$ to be general
Ricci flat, while for the three-form, we choose vanishing background
$\langle C \rangle=0$. For $\mathcal{Y}$ being a toroidal $G_2$
orbifold, the Ricci flatness condition implies $\langle\hat{g}\rangle$
has constant components. In addition, these components should be
constrained in accordance with the orbifold symmetry. Truncating these
11-dimensional fields to our particular singularity, this background
leads to constant seven-dimensional fields $\langle \tau \rangle$ and
$\langle {\ell_I}^{\ul{J}}\rangle$, and vanishing $\langle
A_{\hat{\mu}}^I \rangle$, according to the identifications
\eqref{id1}-\eqref{idn}. Substituting this background into the field
equations for the localised fields gives
\ba
\mathcal{D}F&=&0\, , \\
\mathcal{D}_{\hat{\mu}}F^{a\hat{\mu}\hat{\nu}}&=&e^{\langle\tau\rangle}
     {f^a}_{bc}\phi^{b}_u\hat{\mathcal{D}}^{\hat{\nu}}\phi^{cu}\, ,\\
\hat{\mathcal{D}}_{\hat{\mu}}\hat{\mathcal{D}}^{\hat{\mu}}\phi_{au}&=&
     -\frac{i\sqrt{2}}{2}\epsilon_{uvw}f_{abc}\phi^{bv}
     D^{cw}\, .
\ea
A valid background is thus obtained by setting the genuine
seven-dimensional fields to zero, that is, $\langle A_{\hat{\mu}}^a\rangle=0$,
and $\langle \phi_{au} \rangle =0$. With these fields switched off, the singularity
causes no modification to the background for the bulk fields.

We now discuss supersymmetry of the background. Substitution of our background
into the fermionic supersymmetry transformation laws
\eqref{susyPsi}, \eqref{susyPsi2}, \eqref{susyla} makes every term vanish
except for the $\nabla_M \eta$ term in the variation of the gravitino. Hence,
the existence of precisely one Killing spinor on a $G_2$ space guarantees that
our background is supersymmetric, with $\mathcal{N}=1$ supersymmetry from a
four-dimensional point of view. \\

Let us now discuss the zero modes of these background solutions, both for the
bulk and the localised fields. We begin with the bulk zero modes.
All the orbifold examples discussed in Appendix~\ref{A} restrict the internal metric
to be diagonal (and do not allow any invariant two-forms) and we will focus
on examples of this type in what follows. Hence, the 11-dimensional metric can be written as
\begin{equation} \label{metric1}
\mathrm{d}s^2=\left( \prod_{A=1}^7 R^A \right)^{-1} g_{\mu\nu}dx^\mu dx^\nu +
              \sum_{A=1}^{7}\left(R^A\mathrm{d}x^A\right)^2\, .
\end{equation}
Here the $R^A$ are precisely the seven radii of the underlying
seven-torus. The factor in front of the first part has been
chosen so that $g_{\mu\nu}$ is the four-dimensional metric in the
Einstein frame. There exists a $G_2$ structure $\varphi$, a harmonic
three-form, associated with each Ricci flat metric. For the seven-dimensional
part of the above metric and an appropriate choice of coordinates it is given by
\begin{eqnarray} \label{structure3}
\varphi & = & R^1R^2R^3\mathrm{d}x^1\wedge\mathrm{d}x^2\wedge\mathrm{d}x^3+
R^1R^4R^5\mathrm{d}x^1\wedge\mathrm{d}x^4\wedge\mathrm{d}x^5-
R^1R^6R^7\mathrm{d}x^1\wedge\mathrm{d}x^6\wedge\mathrm{d}x^7 \nonumber \\
 & &+R^2R^4R^6\mathrm{d}x^2\wedge\mathrm{d}x^4\wedge\mathrm{d}x^6 +
R^2R^5R^7\mathrm{d}x^2\wedge\mathrm{d}x^5\wedge\mathrm{d}x^7+
R^3R^4R^7\mathrm{d}x^3\wedge\mathrm{d}x^4\wedge\mathrm{d}x^7 \nonumber \\
 & & -R^3R^5R^6\mathrm{d}x^3\wedge\mathrm{d}x^5\wedge\mathrm{d}x^6.
\end{eqnarray}
It is the coefficients of $\varphi$ that define the metric moduli $a^A$, where
$A=0,\ldots ,6$, in the reduced theory, and these become the real bosonic parts of chiral
superfields. We thus set
\begin{equation} \label{period2}
\left. \begin{array}{cccc}
a^0=R^1R^2R^3, & a^{1}=R^1R^4R^5, & a^{2}=R^1R^6R^7, & a^{3}=R^2R^4R^6, \\
a^{4}=R^2R^5R^7, & a^{5}=R^3R^4R^7, & a^{6}=R^3R^5R^6. &  \, \\
\end{array} \right.
\end{equation}
Since there are no one-forms on a $G_2$ space, and our assumption
above states that there are no two-forms on $\mathcal{Y}$, the three-form field $C$ expands
purely in terms of three-forms, and takes the same form as $\varphi$, thus
\ba \label{Cexp1}
C&=& \nu^0\mathrm{d}x^1\wedge\mathrm{d}x^2\wedge\mathrm{d}x^3+
\nu^{1}\mathrm{d}x^1\wedge\mathrm{d}x^4\wedge\mathrm{d}x^5-
\nu^{2}\mathrm{d}x^1\wedge\mathrm{d}x^6\wedge\mathrm{d}x^7 +
 \nu^{3}\mathrm{d}x^2\wedge\mathrm{d}x^4\wedge\mathrm{d}x^6 \nn \\
 && + \nu^{4}\mathrm{d}x^2\wedge\mathrm{d}x^5\wedge\mathrm{d}x^7+
\nu^{5}\mathrm{d}x^3\wedge\mathrm{d}x^4\wedge\mathrm{d}x^7-
\nu^{6}\mathrm{d}x^3\wedge\mathrm{d}x^5\wedge\mathrm{d}x^6.
\ea
The $\nu^A$ become axions in the reduced theory, and pair up with the metric moduli
to form the superfields
\begin{equation}
\label{TA}
T^A=a^A+i\nu^A.
\end{equation}
In general, not all of the $T^A$ are independent. In fact, a simple
procedure determines which of the $T^A$ are constrained to be
equal. Each generator $\alpha_\tau$ of the orbifold group $\Gamma$
acts by simultaneous rotations in two planes, corresponding to index pairs
$(A,B)$ and $(C,D)$ say. If the order of $\alpha_\tau$ is greater than two, then we
identify $R^A$ with $R^B$ and $R^C$ with $R^D$.  We go through this
process for all generators of $\Gamma$ and then use \eqref{period2} to
determine which of the $a^A$, and hence $T^A$, are equal.

We now discuss a convenient re-labelling of the metric moduli, adapted to the
structure of the singularity. Under the identification of coordinates \eqref{7coordid}, the
metric modulus $a^0$ can be viewed as the volume modulus of the three-torus locus
$\mathcal{T}^3$ of the singularity. The other moduli, meanwhile, are each a product
of one radius of the torus $\mathcal{T}^3$ with two radii of $\mathcal{Y}$ transverse to
the singularity. It is sometimes useful to change the notation for
these moduli to the form $a^{mi}$ where $m=1,2,3$ labels the radius on
$\mathcal{T}^3$ that $a^{mi}$ depends on, and $i=1,2$. Thus
\begin{equation}
\begin{array}{cccccc}
a^{11}=a^1, & a^{12}=a^2, & a^{21}=a^3, & a^{22}=a^4 & a^{31}=a^5,& a^{32}=a^6.
\end{array}
\end{equation}
We will also sometimes make the analogous change of notation for $\nu^A$
and $T^A$.\\

Having listed the bulk moduli, we now turn to the zero modes
associated with the singularity. The decomposition of
seven-dimensional fields works as follows. We take the straightforward
basis $(\mathrm{d}x^m)$ of harmonic one-forms on the three-torus, so
$A^a_{\hat{\mu}}$ simply decomposes into a four-dimensional vector $A^a_{\mu}$ plus the
three scalar fields $A^a_m$ under the reduction. The seven-dimensional
scalars $\phi_{au}$ simply become four-dimensional scalars. Setting
\ba
{b_a}^m&=&-A_{ma}, \\
{\rho_a}^1&=&\sqrt{a^{11}a^{12}}{\phi_a}^3, \\
{\rho_a}^2&=&-\sqrt{a^{21}a^{22}}{\phi_a}^2, \\
{\rho_a}^3&=&\sqrt{a^{31}a^{32}}{\phi_a}^1,
\ea
we can define the complex fields
\begin{equation} \label{cdefn}
{\mathcal{C}_a}^m={\rho_a}^m + i {b_a}^m\; .
\end{equation}
As we will see, the fields ${\mathcal{C}_a}^m$ are indeed the correct
four-dimensional chiral matter superfields. \\

The moduli in the above background solutions are promoted to four-dimensional fields,
as usual, and we will call the corresponding bulk fields $\hat{g}^{(0)}$ and $C^{(0)}$,
in the following. In a pure bulk theory, this would be a standard procedure and
the reduction to four dimension would proceed without further complication.
However, in the presence of localised fields there is a subtlety which we
will now discuss. Allowing the moduli to fluctuate introduces localised
stress energy on the seven-dimensional orbifold plane and this excites the
heavy modes of the theory which we would like to truncate in the reduction.
This phenomenon is well-known from Ho\v rava-Witten theory and can be dealt
with by explicitly integrating out the heavy modes, thereby generating
higher-order corrections to the effective theory~\cite{Lukas:1998ew}.
As we will now argue, in our case these corrections are always of higher
order. More precisely, we will compute the four-dimensional effective theory
up to second order in derivatives and up to order $\kappa_{11}^{4/3}$, relative
to the leading gravitational terms. Let  $\hat{g}^{(1)}$ and $C^{(1)}$
be the first order corrections to the metric and the three-form which originate from
integrating out the localised stress energy on the orbifold plane, so that
we can write for the corrected fields
\ba
\hat{g}^{(\mathrm{B})}&=&\hat{g}^{(0)}+ \kappa_{11}^{4/3} \hat{g}^{(1)}\, ,
\label{backrfields1}\\
C^{(\mathrm{B})}&=&C^{(0)}+\kappa_{11}^{4/3} C^{(1)}\,
.\label{backrfields2} \ea We note that these corrections are already
suppressed by $\kappa_{11}^{4/3}$ relative to the pure background
fields. Therefore, when inserted into the orbifold
action~\eqref{boseaction}, the resulting corrections are of order
$\kappa_{11}^{8/3}$ or higher and will, hence, be neglected.
Inserted into the bulk action, the fields~\eqref{backrfields1} and
\eqref{backrfields2} lead to order $\kappa_{11}^{4/3}$ corrections
which can be written as
\begin{equation}
\delta\mathcal{S}_7=h_7^2\hat{g}_{MN}^{(1)}\left. \frac{\delta \mathcal{S}_{11}}
{\delta \hat{g}_{MN}}\right|_{\hat{g}=\hat{g}^{(0)},\, C=C^{(0)}} +
h_7^2C_{MNP}^{(1)}\left. \frac{\delta \mathcal{S}_{11}}{\delta C_{MNP}}
\right|_{\hat{g}=\hat{g}^{(0)},\, C=C^{(0)}}\, .
\label{dS}
\end{equation}
Let us analyse the properties of the terms contained in this
expression. The functional derivatives in the above expression
vanish for constant moduli fields since the background
configurations $\hat{g}^{(0)}$ and $\hat{C}^{(0)}$ are exact
solutions of the 11-dimensional bulk equations in this case. Hence,
allowing the moduli fields to be functions of the external
coordinates, the functional derivatives must contain at least two
four-dimensional derivatives. All terms in $\hat{g}^{(1)}$ and
$\hat{C}^{(1)}$ with four-dimensional derivatives will, therefore,
generate higher-dimensional derivative terms in four dimensions and
can be neglected. The only terms which are not of this type arise
from the D-term potential and covariant derivatives on the orbifold
plane and they appear within $\hat{g}^{(1)}$. These terms are of
order $\kappa^{4/3}$ and should in principle be kept. However, they
are of fourth order in the matter fields $\mathcal{C}^{am}$ and
contain two four-dimensional derivatives acting on bulk moduli. They
can, therefore, be thought of as corrections to the moduli kinetic
terms. As we will see, the K\"ahler potential of the
four-dimensional theory can be uniquely fixed without knowing these
correction terms explicitly.

\subsection{Calculation of the four-dimensional effective theory}
We will now reduce our theory to four dimensions starting with the
lowest order in the $\kappa$ expansion, that is, with the bulk theory.
The reduction of the bulk theory leads to a well-defined
four-dimensional supergravity theory in its own right. We shall keep
technical discussion to a minimum, and refer the reader to Appendix
\ref{appred} for further details of the method of reduction. The superpotential
and D-term vanish when one reduces 11-dimensional supergravity on a $G_2$
space~\cite{Comp1,Beasley}. Also, we have no gauge fields to consider since our $G_2$
orbifolds do not admit two-forms. Thus we need only specify the K\"ahler potential to
determine the four-dimensional effective theory. To compute this we
use the formula
\begin{equation} \label{Kformula}
K=-\frac{3}{\kappa_4^2}\ln \left( \frac{V}{v_7} \right)\, ,
\end{equation}
 given in Ref.~\cite{Beasley}. Here, $V$ is the volume of the $G_2$ space $\mathcal{Y}$ as
measured by the internal part $g$ of the metric~\eqref{metric1}, and $v_7$ is
a reference volume,
\begin{equation}
V=\int_{\mathcal{Y}}\mathrm{d}^7x\sqrt{ \det g}\, , \hspace{0.5cm}
v_7=\int_{\mathcal{Y}}\mathrm{d}^7x \, .
\end{equation}
Here the four-dimensional Newton constant $\kappa_4$ is related to its
11-dimensional counterpart by
\begin{equation} \label{kappa114}
\kappa_{11}^2=\kappa_4^2v_7\, .
\end{equation}
For the $G_2$ orbifolds considered the volume is proportional to the
product of the seven radii $R^A$ in Eq.~\eqref{metric1}. The precise
form of the K\"ahler potential for 11-dimensional supergravity on
$\mathcal{Y}=\mathcal{T}^7/\Gamma$ is then given by
\begin{equation} \label{kahler1}
K_0=-\frac{1}{\kappa_4^2}\sum_{A=0}^6 \ln \left( T^A + \bar{T}^A
\right)+\frac{7}{\kappa_4^2}\ln 2\, .
\end{equation}
\\

In order to perform the reduction of the seven-dimensional
Yang-Mills theory on the singular locus $\mathcal{T}^3$ to four
dimensions, we need to express the truncated bulk fields
$F_{\hat{\mu}\hat{\nu}}^I$, ${\ell_I}^{\ul{J}}$ and $\tau$ in terms
of the bulk metric moduli $a^A$ and the bulk axions $\nu^A$. We do
this by using the formulae \eqref{metric1}, \eqref{period2},
\eqref{Cexp1} for the 11-dimensional fields in terms of $a^A$ and
$\nu^A$, together with the field identifications
\eqref{id1}-\eqref{idn} between 11-dimensional and seven-dimensional
fields. We find that the only non-vanishing components of
$F_{\hat{\mu}\hat{\nu}}^I$ are some of the mixed components
$F^I_{\mu m}$, and these are given by
\ba
F^3_{\mu 1} =
\frac{1}{2}\left( -\partial_\mu \nu^{11} - \partial_\mu \nu^{12}
\right),
 && F^4_{\mu 1} = \frac{1}{2}\left( -\partial_\mu \nu^{11} + \partial_\mu \nu^{12} \right),
 \nn \\
F^2_{\mu 2} = \frac{1}{2}\left( \partial_\mu \nu^{21} + \partial_\mu
\nu^{22} \right),
 && F^5_{\mu 2} = \frac{1}{2}\left( \partial_\mu \nu^{21} - \partial_\mu \nu^{22} \right),
 \nn \\
F^1_{\mu 3} = \frac{1}{2}\left( -\partial_\mu \nu^{31} -
\partial_\mu \nu^{32} \right),
 && F^6_{\mu 3} = \frac{1}{2}\left( \partial_\mu \nu^{31} - \partial_\mu \nu^{32} \right).
\ea
For the coset matrix $\ell$, which is symmetric, we find the non-zero components
\ba
{\ell_1}^1={\ell_6}^6=\frac{a^{31}+a^{32}}{2\sqrt{a^{31}{a^{32}}}}, && {\ell_1}^6={\ell_6}^1=\frac{a^{31}-a^{32}}{2\sqrt{a^{31}{a^{32}}}}, \nn \\
{\ell_2}^2={\ell_5}^5=\frac{a^{21}+a^{22}}{2\sqrt{a^{21}{a^{22}}}}, && {\ell_2}^5={\ell_5}^2=\frac{-a^{21}+a^{22}}{2\sqrt{a^{21}{a^{22}}}}, \\
{\ell_3}^3={\ell_4}^4=\frac{a^{11}+a^{12}}{2\sqrt{a^{11}{a^{12}}}}, && {\ell_3}^4={\ell_4}^3=\frac{-a^{11}+a^{12}}{2\sqrt{a^{11}{a^{12}}}}. \nn
\ea
Finally, we have the following relation for the orbifold scale factor $\tau$:
\begin{equation}
e^{\tau} = (a^0)^{-2/3}\prod_{m=1}^3\left(a^{m1}a^{m2}\right)^{1/3}.
\end{equation}
We now present the results of our reduction of bosonic terms at the
singularity. We neglect terms of the form $\mathcal{C}^n(\partial
T)^2$, where $n\geq 2$, and thus neglect the back-reaction term
$\delta\mathcal{S}_7$ in Eq.~\eqref{dS} completely. From the
seven-dimensional action $\mathcal{S}_7^{(0)}$ in
Eq.~\eqref{boseaction} we get the following terms, divided up into
scalar kinetic terms, gauge-kinetic terms and scalar potential:
\ba
\mathcal{L}_{4,\mathrm{kin}} &=&
 -\frac{1}{2{\lambda_4}^2}\sqrt{-g}\sum_{m=1}^3\left\{ \frac{1}{a^{m1}a^{m2}}
 \left(\mathcal{D}_\mu\rho_a^m\mathcal{D}^\mu\rho^{am}
 +\mathcal{D}_\mu b_a^m\mathcal{D}^\mu b^{am}\right)\right. \nn \\
 && \hspace{3.0cm}- \frac{1}{3} \sum_{A=0}^6 \frac{1}{a^{m1}a^{m2}a^A}
\partial_\mu a^A\left( \rho_a^m\mathcal{D}^\mu\rho^{am}+b_a^m\mathcal{D}^\mu b^{am}
 \right) \nn \\
&&  \hspace{3.0cm}- \frac{1}{(a^{m1})^2a^{m2}}\rho_{a}^m\left( \partial_\mu
    \nu^{m1}\mathcal{D}^\mu b^{am} + \partial_\mu a^{m1} \mathcal{D}^\mu \rho^{am} \right)
    \nn \\
&& \hspace{3.0cm}\left. -\frac{1}{a^{m1}(a^{m2})^2}\rho_a^m\left(
    \partial_\mu \nu^{m2}\mathcal{D}^\mu b^{am} + \partial_\mu a^{m2}
    \mathcal{D}^\mu \rho^{am} \right) \right\},\label{kinactual} \\
\mathcal{L}_{4,\mathrm{gauge}}&=&-\frac{1}{4\lambda_4^2}\sqrt{-g}\left(a^0
    F_{\mu\nu}^aF^{\mu\nu}_a-\frac{1}{2}\nu^0\epsilon^{\mu\nu\rho\si}
    F^a_{\mu\nu}F_{a\rho\si}\right), \label{gauge}\\
\mathcal{V}&=&\frac{1}{4\lambda_4^2a^0}\sqrt{-g}{f^a}_{bc}f_{ade}
    \sum_{m,n,p=1}^3 \epsilon_{mnp}\frac{1}{a^{n1}a^{n2}a^{p1}a^{p2}}
    \left( \rho^{bn}\rho^{dn}\rho^{cp}\rho^{ep}+\rho^{bn}\rho^{dn}b^{cp}b^{ep}\right. \nn \\
    &&\hspace{7.3cm}\left. +b^{bn}b^{dn}\rho^{cp}\rho^{ep}+b^{bn}b^{dn}b^{cp}b^{ep}\right).
    \label{scalarpot}
\ea
The four-dimensional gauge coupling $\lambda_4$ is related to the seven-dimensional analogue by
\begin{equation} \label{la74}
\lambda_4^{-2}=v_3\lambda_7^{-2}\, ,\qquad v_3=\int_{\mathcal{T}^3}\mathrm{d}^3x\, ,
\end{equation}
where $v_3$ is the reference volume for the three-torus.
Note that the above matter field action is suppressed relative to the gravitational
action by a factor $h_4^2=\kappa_4^2/\lambda_4^2\sim \kappa_{11}^{4/3}$, as mentioned earlier.
\\

\subsection{Finding the superpotential and K\"ahler potential}
The above reduced action must be the bosonic part of a four-dimensional
$\mathcal{N}=1$ supergravity and we would now like to determine the
associated K\"ahler potential and superpotential.
We start by combining the information from the expression \eqref{kahler1}
for the bulk K\"ahler potential $K_0$ descending from 11-dimensional supergravity
with the matter field terms \eqref{kinactual}, \eqref{gauge} and \eqref{scalarpot}
descending from the singularity to obtain the full K\"ahler potential. In general,
one cannot expect the definition~\eqref{TA} of the moduli in terms of
the underlying geometrical fields to remain unchanged in the presence of
additional matter fields. We, therefore, begin by writing the most general
form for the correct superfield $\tilde{T}^A$ in the
presence of matter fields as
\begin{equation}
\ti{T}^A=T^A + F^A\left( T^B, \bar{T}^B, \mathcal{C}_m^a, \bar{\mathcal{C}}_m^a \right)
 \,.\label{Tcorr}
\end{equation}
Analogously, the most general form of the K\"ahler potential in the presence
of matter can be written as
\begin{equation}
K=K_0+K_1\left( T^A, \bar{T}^A, \mathcal{C}_m^a, \bar{\mathcal{C}}_m^a \right)\,. \label{Kcorr}
\end{equation}
Given this general form for the superfields and the K\"ahler potential, we can work
out the resulting matter field kinetic terms by taking second derivatives of $K$ with
respect to $\ti{T}^A$ and $\mathcal{C}_a^m$. Neglecting terms of order $C^n(\partial T)^2$,
as we have done in the reduction to four dimensions, we find
\ba \label{kingeneric}
\mathcal{L}_{4,\mathrm{kin}}&=&-\sqrt{-g}\left\{ \sum_{m,n=1}^3
\frac{\partial^2K_1}{\partial \mathcal{C}_a^m \partial \bar{\mathcal{C}}_b^n}
\mathcal{D}_\mu\mathcal{C}_a^m\mathcal{D}^\mu\bar{\mathcal{C}}_b^n+
\left( 2\sum_{m=1}^3\sum_{A=0}^6\frac{\partial^2K_1}{\partial \mathcal{C}_a^m
 \partial \bar{T}^{A}}\mathcal{D}_\mu\mathcal{C}_a^m\partial^\mu \bar{T}^A+
 \mathrm{c.c.} \right)\right. \nn \\
&& \hspace{1.5cm} + \left. \left( \sum_{A,B=0}^6 \frac{\partial^2K_0}{\partial T^A
   \partial \bar{T}^B}\partial_\mu T^A \partial^\mu \bar{F}^B + \mathrm{c.c.} \right)\right\}\, .
\ea
By matching kinetic terms \eqref{kinactual} from the reduction with
the kinetic terms in the above equation \eqref{kingeneric} we can
uniquely determine the expressions for the superfields $\ti{T}^A$
and the K\"ahler potential. They are given respectively by
\ba
\ti{T}^A &=& T^A -\frac{1}{24\lambda_4^2}\left( T^A+\bar{T}^A \right) \sum_{m=1}^3
 \frac{\mathcal{C}_a^m\bar{\mathcal{C}}^{am}}{(T^{m1}+\bar{T}^{m1})(T^{m2}+\bar{T}^{m2})}\,,\\
K&=& \frac{7}{\kappa_4^2}\ln 2 - \frac{1}{\kappa_4^2}\sum_{A=0}^6
\ln (\ti{T}^A + \bar{\ti{T}}^A ) + \frac{1}{4\lambda_4^2}
\sum_{m=1}^3\frac{(\mathcal{C}_a^m+\bar{\mathcal{C}}_a^m)(\mathcal{C}^{am}+
\bar{\mathcal{C}}^{am})}{(\ti{T}^{m1}+\bar{\ti{T}}^{m1})(\ti{T}^{m2}+\bar{\ti{T}}^{m2})}\,.
\label{1singkahler}
\ea
\\

We now come to the computation of the gauge-kinetic function $f_{ab}$
and the superpotential $W$. The former is straightforward to read off
from the gauge-kinetic part \eqref{gauge} of the reduced action and is given by
\begin{equation}
f_{ab}=\frac{1}{\lambda_4^2}\ti{T}^0\delta_{ab}.
\end{equation}
To find the superpotential, we can compare the scalar potential \eqref{scalarpot} of
the reduced theory to the standard supergravity formula \cite{Wess} for the scalar potential
\begin{equation}
\mathcal{V}=\frac{1}{\kappa_4^4}\sqrt{-g}e^{\kappa_4^2K}\left( K^{X\bar{Y}}\mathcal{D}_X W
\mathcal{D}_{\bar{Y}} \bar{W} - 3 \kappa_4^2 \lvert W \rvert^2\right) +
\sqrt{-g}\frac{1}{2\kappa_4^4}(\mathrm{Re}f)^{-1ab}D_a D_b\, ,
\end{equation}
taking into account the above results for the K\"ahler potential and the gauge
kinetic function. This leads to the superpotential and D-terms
\ba
W &=& \frac{\kappa_4^2}{24\lambda_4^2}f_{abc}\sum_{m,n,p=1}^3
\epsilon_{mnp}\mathcal{C}^{am}\mathcal{C}^{bn}\mathcal{C}^{cp},
\label{1singsuperpot} \\
D_a &=& \frac{2i\kappa_4^2}{\lambda_4^2}f_{abc}\sum_{m=1}^{3} \frac{\mathcal{C}^{bm}\bar{\mathcal{C}}^{cm}}{(\ti{T}^{m1}+\bar{\ti{T}}^{m1}) (\ti{T}^{m2}+\bar{\ti{T}}^{m2})}.
\ea
It can be checked that these D-terms are consistent with the gauging of
the $\mathrm{SU}(N)$ K\"ahler potential isometries, as they should be.
\\

We are now ready to write down our formulae for the quantities that
specify the four-dimensional effective supergravity for M-theory on
$\mathcal{Y}=\mathcal{T}^7/\Gamma$, including the contribution from all singularities.
To do this we simply introduce a sum over the singularities.

Let us present the notation we need to write down these results. We
introduce a label $(\tau,s)$ for each singularity, the index $\tau$
labelling the generators of the orbifold group, and $s$ labelling
the $M_\tau$ fixed points associated with the generator
$\alpha_\tau$. We write $N_\tau$ for the order of the generator
$\alpha_\tau$. For the 16 types of orbifolds, these integer numbers
can be computed from the information provided in Appendix~\ref{A}.
Thus, near a singular point, $\mathcal{Y}$ takes the approximate
form $\mathcal{T}^3_{(\tau,s)}\times
\mathbb{C}^2/\mathbb{Z}_{N_\tau}$, where $\mathcal{T}^3_{(\tau,s)}$
is a three-torus. The matter fields at the singularities we denote
by $(\mathcal{C}^{(\tau,s)})_a^m$ and it is understood that the
index $a$ transforms in the adjoint of $\mathrm{SU}(N_\tau)$. The
gauge couplings depend only on the type of singularity, that is on
the index $\tau$, and are denoted by $\la_{(\tau)}$. M-theory
determines the values of these gauge couplings, and they can be
derived using equations \eqref{couplerel}, \eqref{kappa114} and
\eqref{la74}. We find
\begin{equation}
\la_{(\tau)}^2=\left( 4\pi \right)^{4/3}
\frac{v_7^{1/3}}{v_3^{(\tau)}}\kappa_4^{2/3}\,
\end{equation}
in terms of the reference volumes $v_7$ for $\mathcal{Y}$ and
$v_3^{(\tau)}$ for $\mathcal{T}^3_{(\tau,s)}$.

The respective formulae for the moduli superfields, K\"ahler potential, superpotential and D-term potential are
\begin{equation} \label{superfield}
\ti{T}^A = T^A -\left( T^A+\bar{T}^A \right) \sum_{\tau,s,m}
\frac{1}{24\la_{(\tau)}^2}\frac{(\mathcal{C}^{(\tau,s)})_a^m(\bar{\mathcal{C}}^{(\tau,s)})^{am}}{(T^{B(\tau,m)}+\bar{T}^{B(\tau,m)})(T^{C(\tau,m)}+\bar{T}^{C(\tau,m)})}\,,
\end{equation}
\begin{equation} \label{fullkahler}
K= - \frac{1}{\kappa_4^2}\sum_{A=0}^6 \ln (\ti{T}^A + \bar{\ti{T}}^A ) +
\sum_{\tau,s,m}\frac{1}{4\lambda_{(\tau)}^2}\frac{\left[(\mathcal{C}^{(\tau,s)})_a^m+(\bar{\mathcal{C}}^{(\tau,s)})_a^m\right]\left[(\mathcal{C}^{(\tau,s)})^{am}+(\bar{\mathcal{C}}^{(\tau,s)})^{am}\right]}{(\ti{T}^{B(\tau,m)}+\bar{\ti{T}}^{B(\tau,m)})(\ti{T}^{C(\tau,m)}+\bar{\ti{T}}^{C(\tau,m)})}
+\frac{7}{\kappa_4^2}\ln 2 \,,
\end{equation}
\begin{equation} \label{fullsuperpot}
W = \frac{1}{24}\sum_{\tau,s,m,n,p}\frac{\kappa_4^2}{\lambda_{(\tau)}^2}f_{abc} \epsilon_{mnp}(\mathcal{C}^{(\tau,s)})^{am}(\mathcal{C}^{(\tau,s)})^{bn}(\mathcal{C}^{(\tau,s)})^{cp},
\end{equation}
\begin{equation} \label{fullD}
D_a = 2i\sum_{\tau,s,m}\frac{\kappa_4^2}{\lambda_{(\tau)}^2}f_{abc} \frac{(\mathcal{C}^{(\tau,s)})^{bm}(\bar{\mathcal{C}}^{(\tau,s)})^{cm}}{(\ti{T}^{B(\tau,m)}+\bar{\ti{T}}^{B(\tau,m)})(\ti{T}^{C(\tau,m)}+\bar{\ti{T}}^{C(\tau,m)})}.
\end{equation}
The index functions $B(\tau,m)$, $C(\tau,m)\in \{0,\ldots,6\}$ indicate
by which two of the seven moduli the matter fields are divided by in
equations \eqref{superfield}, \eqref{fullsuperpot} and
\eqref{fullD}. Their values depend only on the generator index $\tau$
and the R-symmetry index $m$. They may be calculated from the formula
\begin{equation}
a^{B(\tau,m)}a^{C(\tau,m)}=\frac{\left( R^m_{(\tau)} \right) ^2
\prod_A R^A}{\prod_n R^n_{(\tau)}}\, ,
\end{equation}
where $R^m_{(\tau)}$ denote the radii of the three-torus
$\mathcal{T}^3_{(\tau,s)}$. The possible
values of the index functions are given in Table~1.
\begin{table}[t]
\begin{center}
\begin{tabular}{|c|c|c|c|c|}
\hline
$\mathrm{Fixed \, directions\, of\, } \alpha_\tau$ &
$(B(\tau,1),C(\tau,1))$ & $(B(\tau,2),C(\tau,2))$ & $(B(\tau,3),C(\tau,3))$ & $h(\tau)$ \\
\hline
(1,2,3) & (1,2) & (3,4) & (5,6) & 0\\
\hline
(1,4,5) & (0,2) & (3,5) & (4,6) & 1\\
\hline
(1,6,7) & (0,1) & (3,6) & (4,5) & 2\\
\hline
(2,4,6) & (0,4) & (1,5) & (2,6) & 3\\
\hline
(2,5,7) & (0,3) & (1,6) & (2,5) & 4\\
\hline
(3,4,7) & (0,6) & (1,3) & (2,4) & 5\\
\hline
(3,5,6) & (0,5) & (1,4) & (2,3) & 6\\
\hline
\end{tabular}
\caption{Values of the index functions $(B(\tau,m),C(\tau,m))$ and
$h(\tau)$ that appear in the superfield definitions,
K\"ahler potential, D-term potential and gauge-kinetic functions.}
\end{center}
\end{table}

There is a universal gauge-kinetic function for each $\mathrm{SU}(N_\tau )$
gauge theory given by
\begin{equation}
f_{(\tau)}=\frac{1}{\la_{(\tau)}^2}\ti{T}^{h(\tau)},
\end{equation}
where $\ti{T}^{h(\tau)}$ is the modulus that corresponds to the volume
of the fixed three-torus $\mathcal{T}^3_{(\tau,s)}$ of the symmetry
$\alpha_\tau$. The value of $h(\tau)$ in terms of the fixed directions
of $\alpha_\tau$ is given in Table~1.
\\

\subsection{Comparison with results for smooth $G_2$ spaces}
As mentioned in the introduction, one can construct a smooth $G_2$
manifold $\mathcal{Y}^{\mathrm{S}}$ by blowing up the singularities of the $G_2$
orbifold $\mathcal{Y}$ \cite{Joyce,Barrett,Lukas}. The moduli K\"ahler potential
for M-theory on this space has been computed in
Refs.~\cite{Barrett,Lukas}. An outline of this calculation, together with the
full result is given in Appendix~\ref{appred}.
Here, we focus on the contribution from a single singularity, which gives
\begin{equation} \label{Ksing}
K =  -\frac{1}{\kappa_4^2}\sum_{A=0}^{6}\ln (T^A+\bar{T}^A) +
     \frac{2}{N c_\Gamma\kappa_4^2}\sum_m\frac{\sum_{i\leq j}
     \left(\sum_{k=i}^{j}(U^{km}+\bar{U}^{km})\right)^2}{(T^{m1}+
     \bar{T}^{m1})(T^{m2}+\bar{T}^{m2})} + \frac{7}{\kappa_4^2}\ln2\, .
\end{equation}
Here, as usual, $T^A$ are the bulk moduli and $U^{im}$ are the blow-up
moduli. As for the formula for the singular manifold, the index $m$
can be thought of as labelling the directions on the three-torus
transverse to the particular blow-up. Each blow-up modulus is
associated with a two-cycle within the blow-up of a given singularity,
and the index $i,j,\ldots =1,\ldots (N-1)$ labels these. Finally,
$c_\Gamma$ is a constant, dependent on the orbifold group.

In computing the K\"ahler potential \eqref{Ksing}, the M-theory action
was taken to be 11-dimensional supergravity, and so the result is
valid when all the moduli, including blow-up moduli, are large
compared to the Planck length. Therefore, the above result for the
K\"ahler potential cannot be applied to the orbifold limit, where
$\mathrm{Re}(U^{im})\rightarrow 0$. However, the corresponding singular result
\eqref{1singkahler} can be used to consider the case of
small blow-up moduli. As discussed in the introduction, the Abelian
components of the matter fields $\mathcal{C}^{im}$ correspond to
moduli associated with the blow-up of the singularity, while the
non-Abelian components correspond to membrane states that are massless
only in the singular limit. Blowing up the singularity is, from this
point of view, described by turning on VEVs for (the real parts of)
the Abelian fields $\mathcal{C}^{im}$ along the D-flat
directions. This, generically, breaks the gauge group $\mathrm{SU}(N)$
to $\mathrm{U}(1)^{N-1}$ and only leaves the $3(N-1)$ matter fields
$\mathcal{C}^{im}$ massless. This massless field content matches
exactly the zero modes of the blown-up geometry. Therefore, by
switching off the non-Abelian components of $\mathcal{C}^{am}$ in
equation \eqref{1singkahler}, one obtains a formula for the moduli
K\"ahler potential for M-theory on $\mathcal{Y}^{\mathrm{S}}$ with small blow-up
moduli. At first glance this is slightly different from the
smooth result \eqref{Ksing} which contains a double-sum over the Abelian
gauge directions. However, we can show that they are actually equivalent.
First, we identify the bulk moduli $T^A$ in \eqref{Ksing} with $\ti{T}^A$ in \eqref{1singkahler}.
One obvious way of introducing a double-sum into the singular result
\eqref{1singkahler} is to introduce a non-standard basis $X_i$ for the Cartan
sub-algebra of $\mathrm{SU}(N)$, which introduces a metric
\begin{equation}
 \label{kij}
 \kappa_{ij}=\mathrm{tr}(X_iX_j)\, .
\end{equation}
Neglecting an overall rescaling of the fields, identification of the smooth
and singular results for $K$ then requires the identity
\begin{equation} \label{identity}
\sum_{i,j}\kappa_{ij}(\mathcal{C}^{im}+\bar{\mathcal{C}}^{im})(\mathcal{C}^{jm}+
\bar{\mathcal{C}}^{jm})=\sum_{i\leq j}\left(\sum_{k=i}^{j}(U^{(k,m)}+\bar{U}^{(k,m)})\right)^2
\end{equation}
to hold. So far we have been assuming the canonical choice $\kappa_{ij}=\delta_{ij}$,
which is realised by the standard generators
\begin{equation}
X_1= \frac{1}{\sqrt{2}}\mathrm{diag}(1,-1,0,\ldots,0)\,, \hspace{0.3cm} X_2= \frac{1}{\sqrt{6}}\mathrm{diag}(1,1,-2,0,\ldots,0)\, , \hspace{0.3cm}  \cdots \, , \nn
\end{equation}
\begin{equation}
X_{N-1}= \frac{1}{\sqrt{N(N-1)}}\mathrm{diag}(1,1,\ldots,1,-(N-1))\,.
\end{equation}
Clearly, the relation~\eqref{identity} cannot be satisfied with a holomorphic
relation between fields for this choice of generators. Instead, from the
RHS of Eq.~\eqref{identity} we need the metric $\kappa_{ij}$ to be
\begin{equation} \label{killingmetric}
\kappa_{ij}= \left\{ \begin{array}{c}
 (N-j)i\, , \hspace{0.3cm} i\leq j\, ,\\
 (N-i)j\, , \hspace{0.3cm} i>j\, .
\end{array} \right.
\end{equation}
From Eq.~\eqref{identity} this particular metric $\kappa_{ij}$ is
positive definite and, hence, there is always a choice of generators
$X_i$ which reproduces this metric via Eq.~\eqref{kij}. For the simplest case $N=2$,
there is only one generator $X_1$ and the above statement becomes trivial.
For the $N=3$ case, a possible choice for the two generators $X_1$ and $X_2$ is
\begin{equation}
X_1=\mathrm{diag}(0,-1,1)\,, \hspace{0.3cm} X_2=\mathrm{diag}(1,0,-1)\,.
\end{equation}
Physically, these specific choices of generators tell us how the Abelian
group $\mathrm{U}(1)^{N-1}$ which appears in the smooth case is embedded into the
$\mathrm{SU}(N)$ group which is present in the singular limit.


\section{Symmetry Breaking and Discussions}

\label{wilflux}

In this section, we will consider more general background configurations than
those discussed in Section 3. We can investigate the effects of such phenomena
as flux vacua and Wilson lines and in addition, we can study how
gauge symmetry is broken in these configurations. In particular, we would like
to examine the explicit symmetry breaking patterns obtained through Wilson
lines, and the effects of $G$- and $F$-flux on our four-dimensional theory.
We will also briefly explore how to express the super-Yang-Mills sector in the
language of $\mathcal{N}=4$ supersymmetry, a rephrasing that will yield new insight into
the structure of our theory close to a singular point in the $G_2$ space.

\subsection{Wilson lines}

We would now like to discuss breaking of the $\mathrm{SU}(N)$ gauge symmetry
through inclusion of Wilson lines in the internal three-torus $\mathcal{T}^3$.
Let us briefly recall the main features of Wilson-line
breaking~\cite{GSW}--\cite{Cvetic}. A Wilson line is a configuration
of the (internal) gauge field  $A^{a}$ with vanishing associated field
strength. For a non-trivial Wilson-line to be possible, the first fundamental group, $\pi_1$,
of the internal space needs to be non-trivial, a condition satisfied in our case,
as $\pi_1(\mathcal{T}^3)=\mathbb{Z}^3$. Practically, a Wilson line around a non-contractible loop $\gamma$ can be
described by
\begin{equation}
U_{\gamma}=P\exp\left(  -i\oint_{\gamma}X_{a}A^{a}{}_{m}dx^{m}\right)
\end{equation}
where $X_{a}$ are the generators of the Lie algebra of the gauge group, $G$.
This expression induces a group homomorphism, $\gamma\rightarrow U_{\gamma}$,
between the fundamental group and the gauge group of our theory.

We can explicitly determine the possible symmetry breaking patterns by
examining particular embeddings (that is, choices of representation) of
the fundamental group into the gauge group. For convenience, we will
focus on gauge groups $\mathrm{SU}(N)$, where $N=2,3,4,6$, since these are
the gauge groups known to arise from explicit constructions of $G_2$
orbifolds~\cite{Barrett}. For example, we may choose a
representation for $\pi _{1}(\mathcal{T}^3)=\mathbb{Z}^{3}$ in the following
way. Let a generic group element of $\mathbb{Z}^{3}$ be given by a
triple of integers (taking addition as the group multiplication),
\begin{equation}
g=\left(  n,m,p\right).
\end{equation}
Then we may embed this in $\mathrm{SU}(4)$ as
\begin{equation}
g=\left(
\begin{array}
[c]{ccc}%
e^{in}{\bf 1}_{2\times2} &  & \\
& 1 & \\
&  & e^{-2 in}%
\end{array}
\right)
\end{equation}
which will clearly break the symmetry to $\mathrm{SU}(2)\times \mathrm{U}(1)\times \mathrm{U}(1)$. There
is, however, a great deal of redundancy in these choices of embedding and the
homomorphisms we define are clearly not unique. For example, we could have
alternately chosen the map so as to take $g$ to an element in the subgroup
$\mathrm{SU}(2)\times \mathrm{SU}(2)\times \mathrm{U}(1)$, say
\begin{equation}
g=\left(
\begin{array}
[c]{ccc}%
(-1)^{n}{\bf 1}_{2\times2} &  & \\
& e^{ im} & \\
&  & e^{- im}%
\end{array}
\right)\, ,
\end{equation}
which would also break $\mathrm{SU}(4)$ to $\mathrm{SU}(2)\times \mathrm{U}(1)\times \mathrm{U}(1)$. A nice
example of the types of reduced symmetry possible with Wilson lines is given
by the following embedding of $\mathbb{Z}^{3}$ into $\mathrm{SU}(6)$~:
\begin{equation}
\left(
\begin{array}
[c]{ccc}%
e^{in}{\bf 1}_{2\times2} &  & \\
& e^{\frac{-2in}{3}}{\bf 1}_{3\times3} & \\
&  & 1
\end{array}
\right)\, .
\end{equation}
This breaks $\mathrm{SU}(6)$ to the subgroup $\mathrm{SU}(3)\times
\mathrm{SU}(2)\times \mathrm{U}(1)\times \mathrm{U}(1)$, which
contains the symmetry group of the Standard Model. (Though even in
this case, our theory does not contain the particle content of the
Standard Model.)

Having given a number of examples, we can now classify in general, which unbroken
subgroups of $\mathrm{SU}(N)$ are possible (using the group-theoretical tools
provided in Ref.~\cite{Groups}). Clearly, the generic unbroken subgroup is
$\mathrm{U}(1)^{N-1}$, however, certain choices of embedding leave a larger symmetry group intact.
These special choices are of particular interest, but we may smoothly deform
from such a choice to a generic solution by varying a parameter in our
embedding. For example, let the mapping of a group element $(n,m,p)$ in
$\mathbb{Z}^{3}$ into $\mathrm{SU}(3)$ be given by
\begin{equation}
g=\left(
\begin{array}
[c]{ccc}%
e^{i\alpha m+ip} &  & \\
& e^{i\alpha n+ip} & \\
&  & e^{-i\alpha(n+m)-2ip}%
\end{array}
\right)
\end{equation}
where the parameter $\alpha$ may be freely varied. For general values of
$\alpha$ this embedding breaks to $\mathrm{U}(1)^{2}$, however for $\alpha=0$ we may
break to the larger group, $\mathrm{SU}(2)\times \mathrm{U}(1)$.

We find the Wilson lines can
break the $\mathrm{SU}(N)$ symmetry group to any subgroup with the rank $N-1$,
with the generic choice being the Cartan algebra itself. In addition, by introducing
a parameter, as in the above $\mathrm{SU}(3)$ example, any of the possible breakings
can be continuously deformed to the generic breaking.
The results for all possible unbroken gauge groups are summarised in Table 2. \begin{table}[ptb]
\begin{center}%
\begin{tabular}
[c]{|c|c|}\hline
Gauge Group & Residual Gauge Groups from Wilson lines\\\hline
$\mathrm{SU}_{2}$ & $\mathrm{U}_1$\\\hline
$\mathrm{SU}_{3}$ & $\mathrm{SU}_{2}\times \mathrm{U}_1$, $\mathrm{U}_1^{2}$\\\hline
$\mathrm{SU}_{4}$ & $\mathrm{SU}_{3}\times \mathrm{U}_1$, $\mathrm{SU}_{2}\times \mathrm{U}_1^{2}$, $\mathrm{SU}_{2}^{2}\times
\mathrm{U}_1$, $\mathrm{U}_1^{3}$\\\hline
$\mathrm{SU}_{6}$ & $\mathrm{SU}_{5}\times \mathrm{U}_1$, $\mathrm{SU}_{4}\times \mathrm{U}_1^{2}$, $\mathrm{SU}_{2}\times
\mathrm{SU}_{3}\times \mathrm{U}_1^{2}$, $\mathrm{SU}_{2}^{2}\times \mathrm{U}_1^{3}$, $\mathrm{SU}_{2}\times \mathrm{U}_1%
^{4}$,\\
& $\mathrm{SU}_{3}\times \mathrm{U}_1^{3}$, $\mathrm{SU}_{2}\times \mathrm{SU}_{4}\times \mathrm{U}_1$, $\mathrm{SU}_{2}%
^{3}\times \mathrm{U}_1^{2}$, $\mathrm{SU}_{3}^{2}\times \mathrm{U}_1$, $\mathrm{U}_1^{5}$\\\hline
\end{tabular}
\end{center}
\caption{The symmetry group reductions in the presence of Wilson lines}%
\end{table}Note that the Cartan subgroups are included as the last entries for
each of the gauge groups. (For other examples of Wilson lines in $G_2$ spaces see, for example, Refs.~\cite{Tamar2},~\cite{Friedmann}.) 

It is worth noting briefly that we can view this symmetry breaking
by Wilson lines in an alternate light in four-dimensions. Rather
than consider a seven-dimensional compactification and Wilson lines,
we could obtain the same results by turning on VEVs for certain
directions of the scalar fields in our four-dimensional
theory\footnote{In fact, the scalars which directly correspond to
Wilson lines in seven dimensions are the axionic, Abelian parts of
the fields ${\cal C}^{am}$.}. For example, if we give generic VEVs
to all the Abelian directions of the scalar fields in
Eq.~(\ref{scalarpot}) we can break the symmetry to a purely Abelian
gauge group. This corresponds to a generic embedding in the Wilson
line picture. Likewise, we can obtain the larger symmetry groups
listed in Table 2 by giving non-generic VEVs to the scalar fields.

\subsection{$\boldsymbol{G}$- and $\boldsymbol{F}$-Flux}

The previous discussion of Wilson lines can be thought of as describing
non-trivial background configurations for which we still maintain the
condition $F=0$ on the field strength. However, to gain a better understanding
of the possible vacua and their effects, we need to consider the
contributions of flux both from bulk and seven-dimensional field strengths.
Let us start with a bulk flux $G_{\mathcal{Y}}$ for the internal
part~\footnote{We do not discuss flux in the external part of $G$.} of the M-theory
four-form field strength $G$. For M-theory compactifications on smooth $G_2$
spaces this was discussed in Ref.~\cite{Beasley}. In our case, all we have
to do is modify this discussion to include possible effects of the
singularities and their associated seven-dimensional gauge theories.
However, inspection of the seven-dimensional gauge field
action~\eqref{boseaction} shows that a non-vanishing internal $G_{\mathcal{Y}}$ will not
generate any additional contributions to the four-dimensional scalar potential,
apart from the ones descending from the bulk. Hence, we can use the
standard formula~\cite{Beasley}
\begin{equation}
W=\frac{1}{4}\int_{\mathcal{Y}}\left(  \frac{1}{2}C+i\varphi\right)  \wedge G_{\mathcal{Y}}\; ,
\label{Wfluxgen}
\end{equation}
where $\mathcal{Y}$ is a general seven-dimensional manifold of
$G_{2}$ holonomy, $C$
is the 3-form of 11-dimensional supergravity and $\varphi$ is the $G_{2}%
$-structure of $X$. For a completely singular $G_2$ space, where the torus moduli
$T^A$ are the only bulk moduli, this formula leads to a flux superpotential
\begin{equation}
 W\sim n_AT^A\; ,
\end{equation}
with flux parameters $n_A$, which has to be added to the ``matter field''
superpotential~\eqref{1singsuperpot}. If some of the singularities are
blown up we also have blow-up moduli $U^{im}$ and the flux superpotential contains
additional terms, thus
\begin{equation}
 W\sim n_AT^A + n_{im}U^{im} \; . \label{Wflux}
\end{equation}
\\

We now turn to a discussion of the seven-dimensional SYM theory at the singuarity.
First, it is natural to ask whether the matter field superpotential~\eqref{1singsuperpot}
can also be obtained from a Gukov-type formula, analogous to Eq.~\eqref{Wfluxgen},
but with an integration over the three-dimensional internal space on which
the gauge theory is compactified. To this end, we begin by defining the complexified
internal gauge field
\begin{equation}
\mathcal{C}_{a}=\rho_{am}\mathrm{d}x^{m}+ib_{am}\mathrm{d}x^{m}\; .
\end{equation}
It is worthwhile to note at this stage, that writing the real parts
of these fields (which are scalar fields in the original
seven-dimensional theory) as forms is, in fact, an example of the
procedure referred to as
`twisting'~\cite{Bershadsky:1995sp,Acharya}. In this particular
case, the twisting amounts to identifying the $R$-symmetry index
($m=1,2,3$) of our original seven-dimensional supergravity with the
tangent space indices of the three-dimensional compact space,
$\mathcal{T}^{3}$. A plausible guess for the Gukov-formula for the
seven-dimensional gauge theory is an expression proportional to the
integral of the complexified Chern-Simons form
\begin{equation}
\omega_{CS}=\left(  \mathcal{F}^{a}\wedge\mathcal{C}_{a}-\frac{1}{3}%
f_{abc}\mathcal{C}^{a}\wedge\mathcal{C}^{b}\wedge\mathcal{C}^{c}\right)
\end{equation}
over the three-dimensional internal space \cite{AcharyaMod}. Here, $\mathcal{F}$ is the
complexified field strength
\begin{equation}
\mathcal{F}^{a}=\mathrm{d}\mathcal{C}^{a}+f^{a}{}_{bc}\mathcal{C}^{b}\wedge
\mathcal{C}^{c}.
\end{equation}
Indeed, if we specialise to the case of vanishing flux, that is $\mathrm{d}\mathcal{C}^a=0$,
our matter field superpotential~\eqref{1singsuperpot} is exactly reproduced by the
formula
\begin{equation}
W=\frac{\kappa_{4}^{2}}{16\lambda_{4}^{2}}\frac{1}{v_{3}}\int_{\mathcal{T}^{3}}%
\omega_{CS}\; . \label{NewGukov}%
\end{equation}

To see that Eq.~\eqref{NewGukov} also correctly incorporates the contributions of
$F$-flux, we can look at the following simple example of an Abelian $F$-flux.
Let the Abelian parts of the gauge field strength, $F^i$, be expanded in a basis of the harmonic
two-forms, $\omega_{m}=\frac{1}{2}\epsilon_{mnp}\mathrm{d}x^n\wedge \mathrm{d}x^p$, on the internal
three-torus $\mathcal{T}^3$, as
\begin{equation}
F^i=f^{im}\omega_m\; ,\label{fluxAnsatz}
\end{equation}
where $f^{im}$ are flux parameters.  Substituting this expression into the seven-dimensional
bosonic action~\eqref{boseaction} and performing a compactification on $\mathcal{T}^3$ we find
a scalar potential which, taking into account the K\"ahler potential~\eqref{1singkahler},
can be reproduced from the superpotential
\begin{equation}
W=\frac{\kappa_{4}^{2}}{8\lambda_{4}^{2}}f_{im}\mathcal{C}^{im}\; . \label{Wfluxabelian}
\end{equation}
This superpotential is exactly reproduced by the Gukov-type
formula~\eqref{NewGukov} which, after substituting the flux
Ansatz~\eqref{fluxAnsatz}, specialises to its Abelian part. Hence, the
formula~\eqref{NewGukov} correctly reproduces the matter field
superpotential as well as the superpotential for Abelian $F$-flux.
The explicit Gukov formula for multiple singularities
is analogous to Eq.~\eqref{NewGukov}, with an additional sum to run over all
singularities as in Eq.~\eqref{fullsuperpot}. We also note that the
$F$-flux superpotential~\eqref{Wfluxabelian} is consistent with the
blow-up part of the $G$-flux superpotential~\eqref{Wflux} when the identification
of the Abelian scalar fields $\mathcal{C}^{im}$ with the blow-up moduli
$U^{im}$ is taken into account.

\subsection{Relation to $\mathcal{N}$=4 Supersymmetric Yang-Mills Theory}

In the previous sections we have explored aspects and modifications of
the four-dimensional $\mathcal{N}=1$ effective theory. We shall now
take a step back and look at the theory without flux, rephrasing it in order to
provide us with several new insights. The M-theory compactification discussed
in the previous sections is clearly $\mathcal{N}=1$ supersymmetric, by
virtue of our choice to compactify on a $G_{2}$ holonomy
space. However, if we neglect the gravity sector (that is, in
particular hold constant the moduli $T^A$) the remaining theory is
$\mathcal{N}=4$ super-Yang-Mills theory, an expected outcome since we
are compactifying the seven-dimensional SYM theory on a three-torus.
We will now make this connection more explicit by matching the
Yang-Mills part of our four-dimensional effective theory with
$\mathcal{N}=4$ SYM theory in its standard form. This connection is of
particular interest since $\mathcal{N}=4$ SYM theory is of central
importance in many current aspects of string theory, particularly in
the context of the AdS/CFT conjecture \cite{Maldacena}. We will begin
with a brief review of the central features of $\mathcal{N}=4$
Yang-Mills theory itself before identifying this structure in our
M-theory compactification.

In addition to a non-Abelian gauge symmetry (given in our case by $\mathrm{SU}(N)$),
the $\mathcal{N}=4$ SYM Lagrangian in
four-dimensions is equipped with an internal $\mathrm{O}(6)\sim \mathrm{SU}(4)$ R-symmetry.
In terms of $\mathcal{N}=1$ language its field content consists of
Yang-Mills multiplets  $(A_{\mu}^{a},\lambda^{a})$, where $a$ is a gauge index,
and a triplet of chiral multiplets $(A^a_m+iB^a_m,\chi^a_m)$ per gauge multiplet,
where $A^a_m$ and $B^a_m$ are real scalars, $\chi^a_m$ are Weyl fermions and
$m,n,\ldots = 1,2,3$. All we require in order to identify fields is the bosonic part of
the $\mathcal{N}=4$ Lagrangian which is given by \cite{Gliozzi,Brink,AdS}
\begin{align}
\mathcal{L}_{\mathcal{N}=4} &  = -\frac{1}{4g^{2}}G_{\mu\nu}^{a}G_{a}^{\mu\nu
}+\frac{\theta}{64\pi^{2}}\epsilon^{\mu\nu\rho\sigma}G_{\mu\nu}^{a}%
G_{a\rho\sigma}-\frac{1}{2}\left(  \mathcal{D}_{\mu}A_{m}^{a}\mathcal{D}^{\mu}A_{a}^{m}%
-\frac{1}{2}\mathcal{D}_{\mu}B_{m}^{a}\mathcal{D}^{\mu}B_{a}^{m}\right)  \nonumber\\
&  \hspace{0.8cm}+\frac{g^{2}}{4}\mathrm{tr}\left( [A_{m},A_{n}][A^{m}%
,A^{n}]+[B_{m},B_{n}][B^{m},B^{n}]+2[A_{m},B_{n}][A^{m},B^{n}]\right)\; .  \label{N4YM}%
\end{align}
With these $\mathcal{N}=4$ definitions in mind, we turn now to the
four-dimensional effective theory~\eqref{kinactual}--\eqref{scalarpot}
derived in the previous sections and consider the case
where the gravity sector is neglected and the geometric moduli are held
constant. This is the situation when we are in the neighbourhood of a singular
point on the $G_{2}$ space and we are neglecting all bulk contributions. By
inspection, we need the following field identifications
\begin{align}
A_{a}^{m} &  =\frac{1}{\lambda_{4}\sqrt{a^{m1}a^{m2}}}\rho_{a}^{m}\, ,\label{Aa}\\
B_{a}^{m} &  =\frac{1}{\lambda_{4}\sqrt{a^{m1}a^{m2}}}b_{a}^{m}\, ,\\
G_{\mu\nu}^{a} &  =F_{\mu\nu}^{a}\; .
\end{align}
The $\mathcal{N}=4$ coupling constants are related to the $\mathcal{N}=1$
constants by%
\begin{equation}
g^{2} =\frac{\lambda_{4}^{2}}{a^{0}}\, ,\qquad
\theta =\frac{8\pi^{2}\nu^{0}}{\lambda_{4}^{2}}.\label{4couplings}%
\end{equation}
With these identifications, Eqs.~(\ref{kinactual})-(\ref{scalarpot}) exactly
reproduce Eq.~(\ref{N4YM}).

We can now consider the Montonen-Olive and S-duality conjecture~\cite{MontonenOlive} in the
context of our theory. This duality acts on the complex coupling
\begin{equation}
\tau\equiv\frac{\theta}{2\pi}-\frac{4\pi i}{g^{2}}%
\end{equation}
by the standard $\mathrm{SL}(2,\mathbb{Z})$ transformation
\begin{equation} \label{mod}
\tau\rightarrow\frac{a\tau+b}{c\tau+d}%
\end{equation}
with $ad-bc=1$ and $a,b,c,d\in$ $\mathbb{Z}$. Note that these transformations
contain in particular $\tau\rightarrow -\frac{1}{\tau}$, an interchange of strong and weak coupling.
Specifically, the S-duality conjecture is the statement that a $\mathcal{N}=4$
Yang-Mills theory with parameter $\tau$ as defined above and gauge group $G$,
is identical to the theory with coupling parameter transformed as in \eqref{mod} and the
dual gauge group, $\widehat{G}$. Note that here ``dual group'' refers to the
Langlands dual group, (which for $G=\mathrm{SU}(N)$, is given by $\widehat
{G}=\mathrm{SU}(N)/\mathbb{Z}_{N}$)~\cite{Dual}.

When we consider the above transformations within the context of our theory,
several interesting features emerge immediately. With the field
identifications in Eq.~(\ref{Aa})--(\ref{4couplings}) we have
\begin{equation}
\tau=-\frac{4\pi i\ti{T}^{0}}{\lambda_{4}^{2}}.
\end{equation}
Therefore, the shift symmetry $\tau\rightarrow\tau+b$ is equivalent
to an axionic shift of $\ti{T}^0$ and
$\tau\rightarrow-\frac{1}{\tau}$ is given by
$\ti{T}^0\rightarrow\frac{1}{\ti{T}^0}$. Since
$\mathrm{Re}(T^0)=a^{0}$ describes the volume of the torus,
$\mathcal{T}^{3}$, S-duality in the present context is really a form
of T-duality.

Bearing in mind this behavior in the Yang-Mills sector, we turn now to the
gravity sector. In a toroidal compactification of M-theory the T-duality
transformation of $\ti{T}^0$ would be part of the U-duality group \cite{U-duality}
and would, therefore, be an exact symmetry. One may speculate that this
is still the case for our compactification on a $G_2$ orbifold and we proceed
to analyse the implications of such an assumption. Examining the
structure of our four-dimensional effective theory~\eqref{kinactual}--\eqref{scalarpot}
we see that the expressions for $K$, $W$ and
$D$ are indeed invariant under axionic shifts of $\ti{T}^0$. However, it is not
so clear what happens for $\ti{T}^0\rightarrow\frac{1}{\ti{T}^0}$.
An initial inspection of Eqs.~\eqref{kinactual}--\eqref{scalarpot}
shows that while the K\"ahler potential changes by%
\begin{equation}
\delta K\sim\ln\left(  \ti{T}^0\bar{\ti{T}}^{0}\right)\; ,
\end{equation}
the kinetic terms and superpotential will remain unchanged.
In order for the whole supergravity theory to be invariant
we need the supergravity function $\mathcal{G}=K+\ln |W|^2$
to be invariant. However as stands, with the $\ti{T}^0$ independent
superpotential~\eqref{1singsuperpot} this is clearly not the case.
One should, however, keep in mind that this superpotential is valid only
in the large radius limit and can, therefore, in principle be subject to
modifications for small $\mathrm{Re}(T^0)$. Such a possible modification which
would make the supergravity function $\mathcal{G}$ invariant and reproduce the
large-radius result~\eqref{1singsuperpot} for large $\mathrm{Re}(T^0)$ is given by
\begin{equation}
W\rightarrow h(\ti{T}^0)W,
\end{equation}
where
\begin{equation}
h(  \ti{T}^0)  = \frac{1}{\eta^{2}(i\ti{T}^0)\left(
j(i\ti{T}^0)-744\right)  ^{1/12}}\,
\end{equation}
and $\eta$ and $j$ are the usual Dedekind $\eta$-function and Jacobi $j$-function.
For large $\mathrm{Re}(T^0)$ the function $h$ can be expanded as
\begin{equation}
 h(\ti{T}^0)=1+2e^{-2\pi\ti{T}^0}+\dots\, .
\end{equation}
Recalling that $\mathrm{Re}(T^0)$ measures the volume of the singular
locus $\mathcal{T}^3$, the above expansion suggests that the function $h$
may arise from membrane instantons wrapping this three-torus.
It would be interesting to verify this by an explicit membrane instanton
calculation along the lines of Ref.~\cite{Harvey:1999as}.

It is well known that there are two dynamical phases in $\mathcal{N}=4$
Yang-Mills theory in four-dimensions \cite{AdS}. A supersymmetric ground
state of the $\mathcal{N}=4$ theory is attained when the full scalar
potential in Eq.~\eqref{N4YM} vanishes. This is equivalent to the condition
\begin{equation}
\left[  Z^{am},Z^{bn}\right]  =0 \label{comm}
\end{equation}
with $Z^{am}=A^{am}+iB^{am}$. There are two classes of solutions to
this equation.  The first, the ``superconformal phase'', corresponds
to the case where $\left\langle Z^{am}\right\rangle =0$ for all
$a,m$. The gauge symmetry is unbroken for this regime, as is the
superconformal symmetry. In the present context, this phase
corresponds to the neighbourhood of a $\mathbb{C}^{2}/\mathbb{Z}_{N}$
singularity in which the full $\mathrm{SU}(N)$ symmetry is present.
As a result of the $\mathcal{N}=4$ supersymmetry in the gauge sector,
this phase will not be destabilised by low-energy gauge dynamics and,
hence, the theory will not be driven away from the orbifold point by
such effects. However, one can also expect a non-perturbative moduli
superpotential from membrane instantons~\cite{Harvey:1999as} whose
precise form for small blow-up cycles is unknown. It would be
interesting to investigate whether such membrane instanton corrections
can stabilise the system at the orbifold point or whether they drive
it away towards the smooth limit.

The second phase, called the ``Coulomb phase'' (or spontaneously
broken phase) corresponds to the flat directions of the potential
where Eq.~\eqref{comm} is satisfied for $\left\langle
  Z^{am}\right\rangle \neq0$.  The dynamics depend upon the amount of
unbroken symmetry. For generic breaking, $\mathrm{SU}(N)$ is reduced
to $\mathrm{U}(1)^{N-1}$. If this breaking is achieved through
non-trivial VEVs in the $A^{am}$ directions it corresponds,
geometrically, to blowing up the singularity in the internal $G_2$
space.


\section{Conclusion and outlook}

In this paper we have constructed, for the first time, the explicit
four-dimensional effective supergravity action for M-theory on a
singular $G_2$ manifold. The class of $G_2$ manifolds for which our
results are valid consists of quotients of seven-tori by discrete
symmetry groups that lead to co-dimension four singularities, around
which the manifold has the structure
$\mathcal{T}^3\times\mathbb{C}^2/\mathbb{Z}_N$.  Breaking the
$\mathrm{SU}(N)$ gauge theory, generically to $\mathrm{U}(1)^{N-1}$,
by assigning VEVs to (the real parts) of the chiral multiplets along
D-flat directions can be interpreted as an effective four-dimensional
description of blowing up the orbifold. We have used this
interpretation to compare our result for the $G_2$ orbifold with its
smooth counterpart obtained in earlier papers~\cite{Lukas,Barrett}. We
find that, subject to choosing the correct embedding of the Abelian
group $\mathrm{U}(1)^{N-1}$ into $\mathrm{SU}(N)$, the results for the
K\"ahler potentials match exactly. This result seems somewhat
surprising given that there does not seem to be a general reason why
the smooth K\"ahler potential should not receive corrections for small
blow-up moduli when the supergravity approximation breaks down. At any
rate, our result allows us to deal with M-theory compactifications
close to and at the singular limit of co-dimension four A-type
singularities. This opens up a whole range of applications, for
example, in the context of wrapped branes and their associated
low-energy physics.

An interesting extension of the work presented here would be to
attempt a more general compactification of M-theory on a $G_2$
manifold whose singular loci are different from $\mathcal{T}^3$. This would
allow a reduction of the $\mathcal{N}=4$ supersymmetry in the gauge
theory sub-sector to $\mathcal{N}=1$, giving rise to richer
infrared gauge dynamics.

In continuing a programme for M-theory phenomenology, we aim
to explicitly include conical singularities into these models,
thereby incorporating charged chiral matter. This problem is
currently under investigation.

\section*{Acknowledgments}
L.~B.~A.~is supported by an NSF Graduate Research Fellowship and
A.~B.~B.~is supported by a PPARC Postgraduate Studentship. A.~L.~is
supported by a Royal Society UK-Japan Joint Project Grant and
M.~Y.~is supported by a JSPS grant for a Japan-UK Joint Research Project.

\section*{Appendix}

\appendix

\section{Review of orbifold based $\boldsymbol{G_2}$ manifolds} \label{A}

\subsection{Construction and Classification of $\boldsymbol{G_2}$ Orbifolds} \label{class}
Let us give a brief definition of what a $G_2$ manifold is so that we
can describe the general idea~\cite{Joyce} of how to construct compact
$G_2$ orbifolds and manifolds. A $G_2$ manifold is a
seven-dimensional Riemannian manifold admitting a globally defined
torsion free $G_2$ structure \cite{Joyce}. A $G_2$ structure
is given by a three-form $\varphi$ which can be written locally as
\begin{eqnarray} \label{canG2}
\varphi & = &
 \mathrm{d}x^1\wedge\mathrm{d}x^2\wedge\mathrm{d}x^3+
 \mathrm{d}x^1\wedge\mathrm{d}x^4\wedge\mathrm{d}x^5-
 \mathrm{d}x^1\wedge\mathrm{d}x^6\wedge\mathrm{d}x^7+
 \mathrm{d}x^2\wedge\mathrm{d}x^4\wedge\mathrm{d}x^6\nonumber \\& &
+\mathrm{d}x^2\wedge\mathrm{d}x^5\wedge\mathrm{d}x^7+
 \mathrm{d}x^3\wedge\mathrm{d}x^4\wedge\mathrm{d}x^7-
 \mathrm{d}x^3\wedge\mathrm{d}x^5\wedge\mathrm{d}x^6\; .
\label{structure}
\end{eqnarray}
The $G_2$ structure is torsion-free if $\varphi$ satisfies
$\mathrm{d}\varphi=\mathrm{d}\ast\varphi=0$. A
$G_2$ manifold has holonomy $G_2$ if and only if its first fundamental
group is finite.
\\

Our starting point for constructing a compact
manifold of $G_2$ holonomy is an arbitrary seven-torus $\mathcal{T}^7$. We then
take the quotient with respect to a finite group $\Gamma$ contained in
$G_2$, such that the resulting orbifold has finite first fundamental
group. We shall refer to $\Gamma$ as the orbifold group. The result is
a $G_2$ manifold with singularities at fixed loci of elements of
$\Gamma$. Smooth $G_2$ manifolds can then be obtained by blowing up the
singularities. Loosely speaking, this involves removing a patch around
the singularity and replacing it with a smooth space of the same
symmetry. Note that, following this construction, the independent
moduli will come from torus radii and from the radii and orientation
of cycles associated with the blow-ups.
\\

We now review, following Ref.~\cite{Barrett}, a classification of
orbifold-based, compact $G_2$ manifolds in terms of the orbifold group
of the manifold. The classification deals with orbifold groups that
lead to a set of co-dimension four singularities, and that act in a
prescribed way on the underlying lattice that defines the
seven-torus. Essentially, each orbifold group element acts by rotating
two orthogonal two-dimensional sub-lattices of the seven-dimensional
lattice. Thus the matrix of an orbifold group element takes the form
\begin{equation}
\boldsymbol{1}_{3\times 3} \oplus \left( \begin{array}{cc}
\cos\theta_1 & -\sin\theta_1 \\
\sin\theta_1 & \cos\theta_1 \end{array} \right)
\oplus
\left( \begin{array}{cc}
\cos\theta_2 & -\sin\theta_2 \\
\sin\theta_2 & \cos\theta_2 \end{array} \right)
\end{equation}
in some coordinate frame. In addition to a rotation, the symmetries
can also contain a translation of the coordinates of the torus. For
such a symmetry to be compatible with a well-defined $G_2$ structure
on the orbifold, it must be possible to embed the
symmetry in $\mathrm{SU}(2)$, and thus we must have $\theta_2=\pm
\theta_1$. Such symmetries are only compatible with a seven-torus if
$\lvert\theta_1\rvert=2\pi/N$ for $N=2,3,4$ or 6.

Using the class of possible generators, one can obtain a class of
discrete symmetry groups from which compact manifolds of $G_2$
holonomy may be constructed. The conditions for some group $\Gamma$ to
be suitable are that there must exist both a seven-dimensional torus
$\mathcal{T}^7$ and $G_2$ structure $\varphi$ preserved by $\Gamma$,
and also that the first fundamental group of the orbifold
$\mathcal{T}^7/\Gamma$ is finite. There are no suitable orbifold
groups with fewer than three generators. The class found in
Ref.~\cite{Barrett} lists all possibilities with precisely
three-generators. Let us present these groups in a form such that they
preserve the particular $G_2$ structure \eqref{canG2}. Define
\ba
R_N&=&\boldsymbol{1}_{3\times 3} \oplus \left( \begin{array}{cc}
\cos(2\pi/N) & -\sin(2\pi/N) \\
\sin(2\pi/N) & \cos(2\pi/N) \end{array} \right)
\oplus
\left( \begin{array}{cc}
\cos(2\pi/N) & -\sin(2\pi/N) \\
\sin(2\pi/N) & \cos(2\pi/N) \end{array} \right) \, ,\\
P_N&=&(1)\oplus \left( \begin{array}{cc}
\cos(2\pi/N) & -\sin(2\pi/N) \\
\sin(2\pi/N) & \cos(2\pi/N) \end{array} \right)
\oplus
\left( \begin{array}{cc}
\cos(2\pi/N) & \sin(2\pi/N) \\
-\sin(2\pi/N) & \cos(2\pi/N) \end{array} \right)
\oplus \boldsymbol{1}_{2\times 2}\, , \\
Q_0 &=&  \mathrm{diag}(-1,1,-1,1,-1,1,-1)\,, \\
Q_1 &=&  \mathrm{diag}(-1,-1,1,1,-1,-1,1)\,, \\
Q_2& :& (x_1,x_2,x_3,x_4,x_5,x_6,x_7) \mapsto (-x_7,x_2,-x_5,x_4,x_3,x_6,x_1)\, , \\
Q_3& :& (x_1,x_2,x_3,x_4,x_5,x_6,x_7) \mapsto (x_3,x_2,-x_1,x_4,x_7,x_6,-x_5)\,, \\
Q_4& :& (x_1,x_2,x_3,x_4,x_5,x_6,x_7) \mapsto (-x_1,-x_2,x_3,-x_4,x_5,x_6,-x_7)\,,\\
Q_5& :& (x_1,x_2,x_3,x_4,x_5,x_6,x_7) \mapsto (x_1,-x_3,x_2,x_5,-x_4,x_6,x_7)\,,\\
Q_6 &:& (x_1,x_2,x_3,x_4,x_5,x_6,x_7) \mapsto (x_1,-x_5,x_4,-x_3,x_2,x_6,x_7)\, .
\ea
Then, one can take as generators $Q_0$ with one of the $P$s and one of the $R$s to obtain the following orbifold groups:
\begin{equation}
\begin{array}{l}
\mathbb{Z}_2 \times \mathbb{Z}_2 \times \mathbb{Z}_2, \\
\mathbb{Z}_2 \ltimes \left( \mathbb{Z}_2 \times \mathbb{Z}_3 \right), \\
\mathbb{Z}_2 \ltimes \left( \mathbb{Z}_2 \times \mathbb{Z}_4 \right), \\
\mathbb{Z}_2 \ltimes \left( \mathbb{Z}_2 \times \mathbb{Z}_6 \right) ,\\
\mathbb{Z}_2 \ltimes \left( \mathbb{Z}_3 \times \mathbb{Z}_3 \right) ,\\
\mathbb{Z}_2 \ltimes \left( \mathbb{Z}_3 \times \mathbb{Z}_6 \right) ,\\
\mathbb{Z}_2 \ltimes \left( \mathbb{Z}_4 \times \mathbb{Z}_4 \right) ,\\
\mathbb{Z}_2 \ltimes \left( \mathbb{Z}_6 \times \mathbb{Z}_6 \right) .\\
\end{array}
\end{equation}
One can take $Q_0$ with $Q_1$ and one of the $R$s to obtain
\begin{equation}
\mathbb{Z}_2^2 \ltimes \mathbb{Z}_N, \; \; N=3, \, 4 \,\, \mathrm{or} \,\, 6.
\end{equation}
Lastly, there are five other, more complicated groups, constructed as follows:
\ba
\mathbb{E}_1 &=:& \langle P_2,\, Q_2,\, R_4\, \lvert\, [P_2,Q_2]=1,\, [P_2,R_4]=1,\, Q_2^2R_4Q_2^2=R_4^{-1}  \rangle, \\
\mathbb{E}_2 &=:& \langle P_2,\, Q_3,\, R_4\, \lvert\,  [P_2,Q_3]=Q_3^2,\, [P_2,R_4]=1,\, Q_3^2R_4Q_3^2=R_4^{-1}  \rangle, \\
\mathbb{E}_3 &=:& \langle Q_4,\, Q_3,\, R_4\, \lvert\, [Q_4,Q_3]=Q_3^2,\, [Q_4,R_4]=R_4^2,\, Q_3^2R_4Q_3^2=R_4^{-1}  \rangle, \\
\mathbb{E}_4 &=:& \langle Q_5,\, Q_3,\, R_4\, \lvert\,  Q_5^2Q_3Q_5^2=Q_3^{-1},\, [Q_5,R_4]=1,\, Q_3^2R_4Q_3^2=R_4^{-1}  \rangle,\\
\mathbb{E}_5 &=:& \langle Q_6,\, Q_3,\, R_4\, \lvert\,  Q_6^2Q_3Q_6^2=Q_3^{-1},\, Q_6^2R_4Q_6^2=R_4^{-1},\, Q_3^2R_4Q_3^2=R_4^{-1}  \rangle.
\ea
 Note
that since we should really be thinking of orbifold group elements as
abstract group elements as opposed to matrices, the commutator is defined here by $[g,h]=g^{-1}h^{-1}gh$.
\\

\subsection{Properties of a class of $\boldsymbol{G_2}$ manifolds} \label{manifolds}

In this sub-section we discuss properties of a general $G_2$
orbifold $\mathcal{Y}=\mathcal{T}^7/\Gamma$ with co-dimension four
fixed points, and a few details of its blown up analogue
$\mathcal{Y}^{\mathrm{S}}$. We assume that points on the torus that
are fixed by one generator of the orbifold group are not fixed by
other generators. Given an orbifold group, this can always be arranged
by incorporating appropriate translations into the generators, and
thus all of the examples of the previous sub-section are
relevant. Under this assumption we have a well-defined blow-up
procedure.
\\

We begin then by discussing the orbifold
$\mathcal{Y}=\mathcal{T}^7/\Gamma$. Let us consider the
homology. There are no one-cycles on a $G_2$ manifold. If $\Gamma$ is
one of the orbifold groups listed in Section \ref{class} then there
are no two-cycles consistent with its symmetries. We will allow for
more general orbifold groups $\Gamma$ in the following and indeed
the main part of the paper as long as they satisfy this condition.
It is then three-cycles that carry the important
information about the geometry of the space. Let us define
three-cycles by setting four of the coordinates $x^A$ to constants
(chosen so there is no intersection with any of the
singularities). The number of these that fall into distinct homology
classes is then given by the number of independent terms in the $G_2$
structure. Let us explain this statement. The $G_2$ structure can
always be chosen so as to contain the seven terms of the standard
$G_2$ structure \eqref{canG2}, with positive coefficients multiplying
them. If we write $\mathcal{R}^A$ for the coefficient in front of the
$(A+1)^{\mathrm{th}}$ term in Eq.~\eqref{canG2}, then by the number of
independent terms we mean the number of $\mathcal{R}^A$s that are not
constrained by the orbifolding. We then write $C^A$ for the cycle
obtained by setting the four coordinates on which the
$(A+1)^{\mathrm{th}}$ term in \eqref{structure} does not depend to
constants, for example,
\begin{equation} \label{bulkcycle}
C^0=\{x^4,x^5,x^6,x^7=\mathrm{const}\}.
\end{equation}
A pair of $C^A$s for which the corresponding $\mathcal{R}^A$s are
independent then belong to distinct homology classes. There is
therefore some subset $\mathcal{C}$ of $\{C^A\}$ that provides a basis for
$H_3(Y,\mathbb{Z})$. We can conclude that the third Betti number $b^3$ is dependent on the orbifold group $\Gamma$ and is in all
cases a positive integer less than or equal to seven. For the class of orbifold groups obtained in Section \ref{class} it takes values as given in Table~3. A description of
the derivation is given in the discussion below on constructing a $G_2$
structure on $\mathcal{Y}$.  \\

\begin{table}
\begin{center}
\begin{tabular}{c|c}
$\Gamma$ & $b^3(\Gamma)$ \\
\hline
$\mathbb{Z}_2\times\mathbb{Z}_2\times\mathbb{Z}_2$ & 7 \\
$\mathbb{Z}_2 \ltimes \left( \mathbb{Z}_2 \times \mathbb{Z}_3 \right)$ & 5 \\
$\mathbb{Z}_2 \ltimes \left( \mathbb{Z}_2 \times \mathbb{Z}_4 \right)$ & 5 \\
$\mathbb{Z}_2 \ltimes \left( \mathbb{Z}_2 \times \mathbb{Z}_6 \right)$ & 5\\
$\mathbb{Z}_2 \ltimes \left( \mathbb{Z}_3 \times \mathbb{Z}_3 \right)$ & 4\\
$\mathbb{Z}_2 \ltimes \left( \mathbb{Z}_3 \times \mathbb{Z}_6 \right)$ & 4\\
$\mathbb{Z}_2 \ltimes \left( \mathbb{Z}_4 \times \mathbb{Z}_4 \right)$ & 4\\
$\mathbb{Z}_2 \ltimes \left( \mathbb{Z}_6 \times \mathbb{Z}_6 \right)$ & 4\\
$\mathbb{Z}_2^2 \ltimes \mathbb{Z}_3$ & 5\\
$\mathbb{Z}_2^2 \ltimes \mathbb{Z}_4$ & 5\\
$\mathbb{Z}_2^2 \ltimes \mathbb{Z}_6$ & 5\\
$\mathbb{E}_1$ & 3\\
$\mathbb{E}_2$ & 3\\
$\mathbb{E}_3$ & 3\\
$\mathbb{E}_4$ & 2\\
$\mathbb{E}_5$ & 1\\
\end{tabular}
\end{center}
\caption{Third Betti numbers of $\mathcal{T}^7/\Gamma$ for different orbifold groups}
\end{table}

We now present the most general Ricci flat metric and
$G_2$ structure on $\mathcal{Y}$. Given there are by assumption
no invariant two-forms, the symmetries restrict the metric to be
diagonal, and thus
\begin{equation} \label{bulkmetric}
\mathrm{d}s^2=\sum_{A=1}^7(R^A\mathrm{d}x^A)^2,
\end{equation}
where the $R^A$ are precisely the seven radii of the torus.

Under a suitable choice of coordinates the $G_2$ structure is obtained from
the flat $G_2$ structure \eqref{canG2} by rescaling $x^A\to
R^Ax^A$, leading to
\begin{eqnarray} \label{structure5}
\varphi & = & R^1R^2R^3\mathrm{d}x^1\wedge\mathrm{d}x^2\wedge\mathrm{d}x^3+R^1R^4R^5\mathrm{d}x^1\wedge\mathrm{d}x^4\wedge\mathrm{d}x^5-R^1R^6R^7\mathrm{d}x^1\wedge\mathrm{d}x^6\wedge\mathrm{d}x^7 \nonumber \\ & &+R^2R^4R^6\mathrm{d}x^2\wedge\mathrm{d}x^4\wedge\mathrm{d}x^6 +R^2R^5R^7\mathrm{d}x^2\wedge\mathrm{d}x^5\wedge\mathrm{d}x^7+R^3R^4R^7\mathrm{d}x^3\wedge\mathrm{d}x^4\wedge\mathrm{d}x^7 \nonumber \\ & & -R^3R^5R^6\mathrm{d}x^3\wedge\mathrm{d}x^5\wedge\mathrm{d}x^6.
\end{eqnarray}
For the orbifolding to preserve the metric some of the $R^A$ must
be set equal to one another. It is straightforward to check that if one of the orbifold symmetries
$\alpha$ involves a rotation in the $(A,B)$ plane by an angle not equal
to $\pi$, then we must set $R^A=R^B$. Following this prescription, it
is easy to find $b^3$ in terms of the orbifold group $\Gamma$.
\\

To complete our description of $\mathcal{Y}$, we briefly mention its singularities. Recall that we are assuming that points on the torus fixed by one generator of the orbifold group are not fixed by other generators. Thus, if we use the index $\tau$ to label the generators of the orbifold group, which each have a certain number $M_\tau$ of fixed points associated with them, then the singularities of $\mathcal{Y}$ can be labelled by the pair $(\tau,s)$, where $s=1,\ldots,M_\tau$. Near a singular point we can then describe $\mathcal{Y}$ by saying it has the approximate form $\mathcal{T}^3_{(\tau,s)}\times\mathbb{C}^2/\mathbb{Z}_{N_{\tau}}$, where $\mathcal{T}^3_{(\tau,s)}$ is a three-torus.
\\

We now move on to discuss the smooth $G_2$ manifold $\mathcal{Y}^{\mathrm{S}}$, constructed by blowing up the singularities of $\mathcal{Y}$. Blowing up a singularity heuristically involves the following. One firstly removes a four-dimensional ball centred around the singularity times the associated fixed three-torus
$\mathcal{T}^3_{(\tau,s)}$. Secondly one replaces the resulting hole by
$\mathcal{T}^3_{(\tau,s)}\times U_{(\tau,s)}$, where $U_{(\tau,s)}$ is the
blow-up of $\mathbb{C}^2/\mathbb{Z}_{N_{\tau}}$, as discussed in Ref.~\cite{Barrett}. The blow-up $U_{(\tau,s)}$ has the same symmetry as $\mathbb{C}^2/\mathbb{Z}_{N_\tau}$ and is derived from a Gibbons-Hawking space (or gravitational multi-instanton) which approaches $\mathbb{C}^2/\mathbb{Z}_{N_\tau}$ asymptotically. Specifically, the central region of $U_{(\tau,s)}$ looks exactly like Gibbons-Hawking space, while the outer region looks exactly like $\mathbb{C}^2/\mathbb{Z}_{N_\tau}$.  Between the central and outer region $U_{(\tau,s)}$ can be thought of as interpolating between Gibbons-Hawking space and flat space. In this way, $\mathcal{Y}^{\mathrm{S}}$ remains smooth as one moves in or out of a blow-up region.

Gibbons-Hawking spaces provide a
generalization of the Eguchi-Hanson space and their different
topological types are labelled by an integer $N$ (where the case $N=2$
corresponds to the Eguchi-Hanson case). While the Eguchi-Hanson space
contains a single two-cycle, the $N^{\rm th}$ Gibbons-Hawking space
contains a sequence $\gamma_1,\ldots,\gamma_{N-1}$ of such cycles at
the ``centre'' of the space.  Only neighbouring cycles $\gamma_i$ and
$\gamma_{i+1}$ intersect and in a single point and, hence, the
intersection matrix $\gamma_i\cdot\gamma_j$ equals the Cartan matrix of
$A_{N-1}$. Asymptotically, the $N^{\rm th}$ Gibbons-Hawking space has
the structure $\mathbb{C}^2/\mathbb{Z}_{N}$.  Accordingly, we take
$N=N_\tau$ when blowing up $\mathbb{C}^2/\mathbb{Z}_{N_{\tau}}$.

The blown-up singularity $\mathcal{T}^3_{(\tau,s)}\times U_{(\tau,s)}$ contributes $N_\tau-1$ two-cycles and $3(N_\tau-1)$ three-cycles to the homology of $\mathcal{Y}^{\mathrm{S}}$. The two-cycles are simply the two-cycles on $U_{(\tau,s)}$, while the three-cycles are formed by taking the Cartesian product of one of these two-cycles with one of the three one-cycles on $\mathcal{T}^3_{(\tau,s)}$.
We can label these three-cycles by $C(\tau,s,i,m)$, where $i$ labels the two-cycles of
$U_{(\tau,s)}$ and $m$ labels
the direction on $\mathcal{T}^3_{(\tau,s)}$. We deduce the following formula for the third Betti number of $\mathcal{Y}^{\mathrm{S}}$:
\begin{equation}
b^3(\mathcal{Y}^{\mathrm{S}})=b^3(\Gamma)+\sum_\tau M_\tau \cdot 3(N_\tau-1).
\end{equation}

On one of the blow-ups $\mathcal{T}^3\times U$ (for convenience we suppress $\tau$ and $s$ indices) we use coordinates $\xi^m$ on $\mathcal{T}^3$ and complex coordinates $z^p$, $p=1,2$ on $U$. We write $R^m$ to
denote the three radii of $\mathcal{T}^3$, which will be the three $R^A$ in the
directions fixed by $\alpha$. The $G_2$ structure can be written as
\begin{equation} \label{structure4}
\varphi=\sum_m\omega^m\left(z^p,\boldsymbol{b}_1,\ldots,\boldsymbol{b}_{N_\tau}\right)\wedge R^m\mathrm{d}\xi^m-R^1R^2R^3\mathrm{d}\xi^1\wedge\mathrm{d}\xi^2\wedge\mathrm{d}\xi^3.
\end{equation}
Here the $\boldsymbol{b}_i$ are a
set of three-vectors, which parameterize the size of the two-cycles within the Gibbons-Hawking space, and also their orientation with respect to the bulk. The $\omega^m$ are a triplet
of two-forms that constitute a ``nearly'' hyperk\"ahler structure on
$U$. This $G_2$ structure makes a slight deviation from being torsion free in the region in which the space $U$ is interpolating between Gibbons-Hawking space and flat space. However, for sufficiently small blow-up moduli and a smooth and slowly varying interpolation this deviation is small \cite{Barrett,Lukas}. Consequently, this $G_2$ structure can reliably be used in M-theory calculations that work to leading non-trivial order in the blow-up moduli.

The metric corresponding to the $G_2$ structure \eqref{structure4} takes the form
\begin{equation}
\mathrm{d}s^2=\mathcal{G}_0\mathrm{d}\boldsymbol{\zeta}^2+\sum_{m=1}^{3}\mathcal{G}_m(\mathrm{d}\xi^m)^2,
\end{equation}
where $\mathrm{d}\boldsymbol{\zeta}$ is the line element on $U$, and the $\mathcal{G}$s
are conformal factors whose product is equal to 1.


\section{Reduction of M-theory on a $G_2$ manifold} \label{appred}
In this section we briefly review the reduction of M-theory on a smooth $G_2$ manifold, and then give a formula for the K\"ahler potential for M-theory on the orbifold based $G_2$ manifolds we have been discussing.
\\

On a $G_2$ manifold $X$, there is an isomorphism between torsion-free $G_2$ structures and Ricci flat metrics (see for example Ref. \cite{Lukas}). Thus Ricci-flat deformations of the metric can be described by the torsion-free deformations of the $G_2$ structure and, hence, by the third cohomology $H^3(X,\mathbb{R})$. Consequently, the number of independent metric moduli is given by the third Betti number $b^3(X)$. To define these moduli explicitly, we
introduce to $X$ an integral basis $\{C^A\}$ of three-cycles, and a dual basis $\{\Phi_A\}$ of harmonic three forms satisfying
\begin{equation} \label{dual}
\int_{C^A}\Phi_B=\delta^A_B,
\end{equation}
where $A,B,\ldots=1,\ldots,b^3(X)$.
 We can then expand the torsion-free $G_2$ structure $\varphi$ as
\begin{equation}
\varphi=\sum_Aa^A\Phi_A.
\end{equation}
Then, by equation \eqref{dual}, the $a^A$ can be computed in terms of certain underlying geometrical parameters by performing the period integrals
\begin{equation}
a^A=\int_{C^A}\varphi.
\end{equation}
\\
\begin{table}[t]
\begin{center}
\begin{tabular}{|c|c|c|c|}
\hline
$\mathrm{Fixed \, directions\, of\, } \alpha_\tau$
&$(A(\tau,1),B(\tau,1))$ & $(A(\tau,2),B(\tau,2))$ &$ (A(\tau,3),B(\tau,3))$ \\
\hline
(1,2,3) & (1,2) & (3,4) & (5,6) \\
\hline
(1,4,5) & (0,2) & (3,5) & (4,6) \\
\hline
(1,6,7) & (0,1) & (3,6) & (4,5) \\
\hline
(2,4,6) & (0,4) & (1,5) & (2,6) \\
\hline
(2,5,7) & (0,3) & (1,6) & (2,5) \\
\hline
(3,4,7) & (0,6) & (1,3) & (2,4) \\
\hline
(3,5,6) & (0,5) & (1,4) & (2,3) \\
\hline
\end{tabular}
\caption{Values of the index functions $(A(\tau,m),B(\tau,m))$ that
appear in the K\"ahler potential.}
\end{center}
\end{table}

Let us also introduce an integral basis $\{D^I\}$ of two-cycles, where $I,J,\ldots=1,\ldots b^2(X)$, and a dual basis $\{\omega_I\}$ of two-forms satisfying
\begin{equation}
\int_{D^I}\omega_J=\delta^I_J\,.
\end{equation}
Then, the three-form field $C$ of 11-dimensional supergravity can be expanded in terms of the basis $\{\Phi_A\}$ and $\{\omega_I\}$ as
\begin{equation} \label{Cexp}
C=\nu^A\Phi_A+A^I\omega_I.
\end{equation}
The expansion coefficients $\nu^A$ represent $b^3(X)$ axionic fields in the four-dimensional effective theory, while the Abelian gauge fields $A^I$, with field strengths $F^I$, are part of $b^2(X)$ Abelian vector multiplets. The $\nu^A$ pair up with the metric moduli $a^A$ to form the bosonic parts of $b^3(X)$ four-dimensional chiral superfields
\begin{equation}
T^A=a^A+i\nu^A.
\end{equation}

Using general properties of $G_2$ manifolds, it was shown in Ref. \cite{Beasley} that the K\"ahler metric descends from the K\"ahler potential
\begin{equation} \label{Kapp}
K=-\frac{3}{\kappa_4^2}\ln \left( \frac{V}{v_7} \right)\, ,
\end{equation}
where $V$ is the volume of $X$ as measured by the dynamical Ricci flat metric $g$, and $v_7$ is a reference volume, thus
\begin{equation}
V=\int_{X}\mathrm{d}^7x\sqrt{ \det g}\, , \hspace{0.5cm} v_7=\int_{X}\mathrm{d}^7x\sqrt{\det g_0}\, .
\end{equation}
The four-dimensional Newton constant $\kappa_4$ is related to the 11-dimensional version by
\begin{equation}
\kappa_{11}^2=\kappa_4^2v_7\, .
\end{equation}

Reduction of the Chern-Simons term of 11-dimensional supergravity by inserting the gauge field part of \eqref{Cexp} leads to the four-dimensional term \cite{Beasley}
\begin{equation}
\int_{\mathcal{M}_4}c_{AJK}\nu^AF^J\wedge F^K\,,
\end{equation}
where the coefficients $c_{AJK}$ are given by
\begin{equation}
c_{AJK}\sim \int_{X} \Phi_A \wedge \omega_J \wedge \omega_K.
\end{equation}
This implies that the gauge-kinetic function $f_{JK}$, which couples $F^J$ and $F^K$, takes the form
\begin{equation} \label{gkfapp}
f_{JK} \sim \sum_AT^Ac_{AJK}.
\end{equation}
For the case of a non-flux background, the superpotential and D-term potential vanish and thus the K\"ahler potential \eqref{Kapp} and gauge-kinetic function \eqref{gkfapp} fully specify the four-dimensional effective theory.
\\

Let us now present the K\"ahler potential for M-theory on the orbifold based $G_2$ manifold $\mathcal{Y}^{\mathrm{S}}$ described in the previous section. First we give the bulk metric moduli $a^A$ given by
\begin{equation}
a^A=\int_{C^A}\varphi,
\end{equation}
where the $\{C^A\}$ are those cycles described in Section \ref{manifolds}; for example $C^0$ is given by \eqref{bulkcycle}. These simply evaluate to
\begin{equation}
\left. \begin{array}{cccc}
a^0=R^1R^2R^3, & a^1=R^1R^4R^5, & a^2=R^1R^6R^7, & a^3=R^2R^4R^6, \\
a^4=R^2R^5R^7, & a^5=R^3R^4R^7, & a^6=R^3R^5R^6. &  \, \\
\end{array} \right.
\end{equation}
Then the blow-up moduli, which are defined by
\begin{equation}
A(\tau,s,i,m)=\int_{C(\tau,s,i,m)}\varphi
\end{equation}
take the form
\begin{equation}
A(\tau,s,i,m)\sim R^m_{(\tau)}(b_{(\tau,s,i,m)}-b_{(\tau,s,i+1,m)})
\end{equation}
where $R^m_{(\tau)}$ denote the three radii of
$\mathcal{T}^3_{(\tau,s)}$, and $b_{(\tau,s,i,m)}$ are the parameters for the two-cycles within the blow-ups, consistent with the notation of equation
\eqref{structure4}. We denote the superfields associated with the bulk by $T^A$ and those associated with blow-ups by $U^{(\tau,s,i,m)}$, so
\begin{equation}
\mathrm{Re}(T^A)=a^A, \; \; \; \mathrm{Re}(U^{(\tau,s,i,m)})=A(\tau,s,i,m).
\end{equation}
The K\"ahler potential is given by the following formula, to leading
non-trivial order in the blow-up moduli, (taking a trivial reference
metric $g_0=1$):
\begin{equation} \label{K}
K =  -\frac{1}{\kappa_4^2}\sum_{A=0}^{6}\ln (T^A+\bar{T}^A) +\frac{2}{c_\Gamma\kappa_4^2}\sum_{s,\tau,m}\frac{1}{N_\tau}\frac{\sum_{i<j}\left(\sum_{k=i}^{j-1}(U^{(\tau,s,k,m)}+\bar{U}^{(\tau,s,k,m)})\right)^2}{(T^{A(\tau,m)}+\bar{T}^{A(\tau,m)})(T^{B(\tau,m)}+\bar{T}^{B(\tau,m)})} + \frac{7}{\kappa_4^2}\ln2.
\end{equation}
The index functions $A(\tau,m)$, $B(\tau,m)\in\{1,\ldots,7\}$ indicate
by which two of the seven bulk moduli $T^A$ the blow up moduli
$U^{(\tau,s,i,m)}$ are divided in the K\"ahler potential \eqref{K}. Their values depend only on the generator index $\tau$ and the
orientation index $m$. They may be calculated from the formula
\begin{equation}
a^{A(\tau,m)}a^{B(\tau,m)}=\frac{\left(R^m_{(\tau)}\right)^2\prod_AR^A}{\prod_bR^b_{(\tau)}}.
\end{equation}
The $\tau$ dependence is only through the fixed directions of the generator $\alpha_\tau$ and the possible values of the index functions are given in Table~4. We remind the reader that in many cases some of the $T^A$s are identical to each other and should be thought of as the same field. One follows the prescription given in Section \ref{manifolds} to determine which of these are identical. For the orbifold groups $\Gamma$ listed in Section \ref{class}, the number $b^3(\Gamma)$ of distinct $T^A$ is given in Table~3. Finally, $c_\Gamma$ is just a constant factor that depends on the orbifold group.


\end{document}